\documentclass[aip,jcp,preprint,amsmath,amssymb, floatfix]{revtex4-1}

\usepackage{amsmath}
\usepackage{bm}
\usepackage{cancel} 
\usepackage{graphicx}

\usepackage{hyperref}
\usepackage{ifpdf}
\usepackage{rotating}
\usepackage{alltt}
\usepackage{float}
\usepackage{fullpage}
\usepackage{color}
\usepackage[usenames,dvipsnames]{xcolor}

\usepackage{xcolor}

\definecolor{listinggray}{gray}{0.9}
\definecolor{lbcolor}{rgb}{0.9,0.9,0.9}
\definecolor{comment}{rgb}{0.133,0.545,0.133}
\definecolor{function}{rgb}{0.8,0.1,0.1}
\definecolor{keyword}{rgb}{0,0,1}

\usepackage{listings}
\DeclareMathOperator*{\define}{\equiv}

\newcommand{\eq}[1]{ Eq.\ (\ref{#1})}

\DeclareMathOperator*{\Stresstype}{\boldsymbol{\Pi}}

\DeclareMathOperator*{\SurfaceTensiontype}{\gamma}

\DeclareMathOperator*{\StressSurfscalar}{\Stresstype\limits^{\scriptscriptstyle{Surf}}}
\DeclareMathOperator*{\StressSurf}{\boldsymbol{\Stresstype\limits^{\scriptscriptstyle{Surf}}}}
\DeclareMathOperator*{\PressureVA}{\boldsymbol{\Stresstype\limits^{\scriptscriptstyle{V\!A}}}}
\DeclareMathOperator*{\PressureVAscalar}{\Stresstype\limits^{\scriptscriptstyle{V\!A}}}

\newcommand{\insurf}[1]{{\xi}^{\textit{#1}}}

\newcommand{\insurfi}[1]{{\xi_{i}\!}^{\textit{#1}}}

\newcommand{\insurfl}[1]{\xi_{\lambda}^{\textit{#1}}}

\begin{document}

\title{Hydrodynamics Across a Fluctuating Interface}

\author{Edward R. Smith}
\email[]{edward.smith@brunel.ac.uk}
\affiliation{Department of Mechanical and Aerospace Engineering, Brunel University London}

\author{Carlos Braga}
\email[]{c.braga@qmul.ac.uk}
\affiliation{School of Engineering and Materials Science, Queen Mary University of London}

\begin{abstract}
\section*{Abstract}

Understanding what happens inside the rippling and dancing surface of a liquid remains one of the great challenges of fluid dynamics.
Using molecular dynamics (MD) we can pick apart the interface structure and understand surface tension.
In this work we derive an exact mechanical formulation of hydrodynamics for a liquid-vapour interface using a control volume which moves with the surface.
This mathematical framework provides the local definition of hydrodynamic fluxes at any point on the surface. 
These are represented not only by the flux of molecules and intermolecular interactions acting across the surface, but also as a result of the instantaneous local curvature and movement of the surface itself.
By explicitly including the surface dynamics in the equations of motion, we demonstrate an exact balance between kinetic and configurational pressure normal to the surface.
The hydrodynamic analysis makes no assumptions regarding the probability distribution function, so is valid for any system arbitrarily far from thermodynamic equilibrium.
The presented equations provide a theoretical basis for the study of time-evolving interface phenomena such as bubble nucleation, droplet dynamics and liquid-vapour instabilities.

\end{abstract}

\maketitle

% !TEX root =  main.tex

% -----------------------------------------------------------------------------
\section{Introduction}

%--------------

The liquid-vapour interface stands out as one of the great modelling challenges in engineering and the physical sciences.
Central to this challenge is the observation that, while bulk fluids and gases are well described by continuum models, this approach breaks down at the interface, where large changes in the physical properties, localised to a very small region, invalidate any continuum assumptions.
% While on average fluids and gases are well described by models of a homogeneous medium, this breaks down at the interface, where a large change localised to a very small region invalidate continuum assumptions.
Molecular dynamics (MD) is an ideal model to study a problem of this type, modelling the liquid-vapour coexistence with no more than the solution of Newton's law for a system of molecules.

However, using a discrete MD model introduces a new challenge, translation of the information at the molecular level to relevant continuum hydrodynamic properties.
Thermal fluctuations, characteristic of the dynamics at the atomistic scale, are present at the interface, blurring any property of interest. 
Capillary wave theory (CWT)~\citep{Evans1979aa, Rowlinson2002aa} provides a framework to describe these fluctuations, by representing the instantaneous shape of the fluid surface as an intrinsic surface in parametric form.
% $z=\xi\bb{x,y,t}$, with an effective Hamiltonian $H\bbs{\xi}=\gamma\,A\bbs{\xi}$ proportional to the area of the corrugated interface, $A\bbs{\xi}$, with $\gamma$ the surface tension.
Thermodynamic profiles across the interface represent the convolution of an assumed sharp intrinsic profile with the Gaussian distribution characterising the height of the intrinsic surface.

%--------------

A mechanical route is also possible, based on instantaneous time evolving equations with no need for thermodynamics averaging, equivalent to the Newtonian mechanics which underpin molecular dynamics.
Using the sharp intrinsic surface directly as a moving reference frame, mechanical equations can be obtained from the molecular data with no thermodynamic blurring.
The most important quantity describing the mechanical properties of the interfacial region is the pressure tensor. 
For a homogeneous system of particles with periodic boundaries, a single virial pressure tensor can be defined for the whole system \citep{parker}.
In inhomogeneous systems such as the liquid-vapour interface we require a local definition.
The seminal work of \citet{Irving_Kirkwood} provides a localisation to give the pressure at a point in space using the Dirac delta function.
There are two considerations with the \citet{Irving_Kirkwood} stress, the practical and mathematical consequences of the Dirac delta function in an MD simulation and the non-uniqueness in the definition of the pressure tensor \citep{Schofield_Henderson}.

We consider the problematic Dirac delta functions first, a direct consequence of the mathematical idealisation of a continuum.
Several approaches have been applied to solve this, including mollifying the Delta function \citep{Noll, Admal_Tadmor, Murdoch}, integrating to get "Volume Average" (VA) stress \citep{Lutsko, Cormier_et_al} or reformulating in terms of the pressure over a surface \citep{Tsai, Todd_et_al_95, Han_Lee}, known as the method of planes (MOP) pressure.
Of these approaches, the pressure over the surface has the advantage of being conceptually the simplest form, namely the force acting over the surface divided by the surface area $\boldsymbol{\StressSurf} = \boldsymbol{F}/A$.
It is also the only form to give exactly conservative equations at every MD timestep, shown through the concept of a control volume \citep{Smith_et_al12}.
The control volume equations are important in fluid dynamics, providing the basis for the so-called conservative finite volume approach in computational fluid dynamics (CFD) \citep{Hirsch}.
By expressing a weakened statement of the equations of motion, conserved in an average sense over an arbitrary volume, they no longer demand a continuous field.
This makes the control volume approach ideally suited to molecular dynamics, expressing everything as an average over a volume.
This solves the problem of the Dirac delta functions by expressing the \citet{Irving_Kirkwood} equations in terms of integrated quantities inside a volume whose evolution is exactly equal to the sum of pressure, $\boldsymbol{\StressSurf}$, over the bounding surface.

Next we consider the non-uniqueness, argued to make mechanical definitions unreliable \citep{Malijevsk__2012, C9CP02890K}.
The ambiguity in pressure tensor consists of at least three possible components: $i)$  the inclusion of kinetic terms \citep{Zhou}, $ii)$ the intermolecular interaction path \citep{Schofield_Henderson} and $iii)$ the choice of measuring reference frame \citep{Admal_Tadmor}.
By providing an exact mechanical link between surface pressure and time evolution of momentum, the conservative nature of the control volume formulation helps to sidestep some issues associated with the non-uniqueness of the pressure tensor.
% and $iv)$ the choice of guage \citep{}.
The inclusion of kinetic terms $i)$ in the pressure has been shown to be essential so surface pressure is equal to the momentum change inside the control volume \citep{Smith_et_al12}, a result extended in this work to include movement of the surface itself.
The second source of non-uniqueness $ii)$, stems from the infinite number of possible interaction paths between atoms.
%, a result of classical MD reducing the complexity of the quantum interaction to a potential acting between point atoms.
In this work, a linear path is used, sometimes called the Irving Kirkwood contour, which is consistent with the definition of "impressed force" used by Newton \citep{Newton, Newton_translated}.
Another common interaction path is the Harasima contour \citep{Harasima1958aa}, known to give unphysical results for spherical coordinates \citep{PhysRevE.66.011203} and require adjustment for cylindrical coordinates \citep{10.1021/acs.jctc.0c00607}.
The Harasima contour also combines path with reference frame, linking issues $ii)$ and $iii)$, as the tangential contour follows the interface.
The Harasima contour would follow the intrinsic interface in this work with varying normals at each point, and may be expected to experience similar problems to the spherical and cylindrical cases.
It would also be difficult and computationally expensive to implement as an integral following the intrinsic surface.

The measuring reference frame, $iii)$, used in this work is a closed bounding surface of a control volume.
A direct consequence of using MOP style pressure defined over this bounding surface, $\oint \boldsymbol{\StressSurf} \cdot d\textbf{S} = \sum_I^{Surfaces} \boldsymbol{F_I}/A_I$, is it simply adds forces between particles over a surface, $F_I = \sum_{i,j}^{N_{I surf}} \boldsymbol{f}_{ij}$, so we retain the exact balance of forces to momentum change from Newtons laws.
As noted by \citet{Schofield_Henderson}, individual pressure is non-unique but the total pressure over the closed surfaces of a volume is invariant to choice of contour.
The result here extends this idea away from equilibrium, mathematically linking the invariant pressure over all surfaces to momentum change inside the volume, $d/dt\sum_{i}^{N_{Vol}} m_i \boldsymbol{v}_i$, and including deformation of the volume with movement of the intrinsic interface.
%The use of a straight path is consistent with Newtons laws for the individual particles and the resulting pressure, although not-unique, is the one that is consistent with the momentum change inside the volume.
It is hoped that this exact link between the invariant closed surface pressure and the resulting particle motion on an interface shown in this work could provide a way forward in addressing the arbitrary nature of the pressure tensor and concerns that it renders results meaningless \citep{Malijevsk__2012, C9CP02890K, C9CP04289J}.
%Instead, the more general instantaneous mechanical $\oint_S \boldsymbol{\Pi} \cdot d\textbf{S}= d/dt \int_V \rho \boldsymbol{u}$ ensures.
This treatment only considers two body interactions in this work, for three body interaction \citep{todd_daivis_2017} or long-range contributions \citep{10.1021/acs.jctc.0c00607} ensuring conservation would require more care.

%The guage invariancd of $iv)$ means a symmetric divergenceless tensor can be added to the pressure without changing the equations of motion.
%In principle, physical arguments for pressure allow us to suggest this is simply a case of choosing a scalar guage pressure for the entire global fluid which can be added to the tensor diagonal.

In addition to conservation, we take advantage of an additional benefit of the control volume approach, the ability to define them with arbitrary shape.
We use the function form of the liquid-vapour interface, $\xi$, and define the face of our control volumes so the edge of our averaging grid follows the 2D surface.
To define this surface, we apply the intrinsic interface approach first proposed by \citet{Chacon2003aa}, fitting a set of Fourier components to the outermost molecules in a liquid cluster.
A range of other approaches to define an interface have been proposed \citep{Partay2008ab, Willard2010aa, Jorge2010ab, Sega2013aa}, and the surface pressure presented in this work can, in principle, be applied to any other surface, not just the \citet{Chacon2003aa} functional form.

%--------------

An assumption of CWT is that the density and pressures profiles are uncorrelated with the intrinsic surface from which the profiles are obtained.
However, in order to balance the control volume equations of motion, it will be shown that it is necessary to explicitly include the pressure change due to the surface time evolution.
The surface itself becomes part of the equations of motion, an addition which is demonstrated here to be essential in obtaining the correct form of the equations of motion.
Constant normal pressure over a liquid-vapour interface is expected \citep{Berry1971aa, Sega2016ab} and has been shown in the literature with a fixed reference frame \citep{10.1080/00268978300100971}.
Near solid-liquid interfaces a constant normal pressure must exist for momentum balance, which the IK1 form of pressure fails to show \citep{10.1063/1.1288390}, an observation which motivates use of the VA and MOP formulations \citep{Todd_et_al_95, Heyes_et_al}.
Applying the same approach to the liquid-vapour interface, the normal component of the intrinsic pressure does not show momentum balance even using the VA form of pressure \citep{Braga_et_al18}.% even with the a direct consequence of not including instantaneous surface movement.
We show that the surface pressure equations, derived in this work gives, give the expected constant normal pressure.
This flat profile requires the pressure to include both the instantaneous normal to the intrinsic surface at every time and a term for the surface movement itself.
The resulting equations instantaneously include all time evolving terms balancing forces and fluxes at each timestep, meaning that the commonly assumed thermodynamic or average equilibrium condition $\boldsymbol{\nabla} \cdot \boldsymbol{\Pi} = 0$  \citep{Rowlinson93, Malijevsk__2012, 10.1080/00268978300100971} is neither required nor valid.
As a result, the presented equations will allow an exploration of the instantaneous hydrodynamics of the interface, known to persist even with a very small number of molecules moving over very short timescales \citep{Delgado-Buscalioni2008aa}.

The structure of this manuscript is as follows, in the theory section \ref{sec:Theory} the mathematical form of the control volume is derived to give the expression for surface fluxes on a volume moving with the intrinsic interface. The next section outlines the details of implementation \ref{sec:Implementation}, giving an overview of how to actually use these expressions in molecular simulations and the details of the setup. The result and discussion are included in section \ref{sec:Results and Discussion} before concluding remarks in section \ref{sec:Conclusions}.

			% section{Introduction}
% !TEX root =  main.tex

% -----------------------------------------------------------------------------
\section{Theory}
\label{sec:Theory}

% To do equation checklist
% \begin{enumerate}
% \item Ensure all signs correct 
% \item Check super and subscripts on surfaces
% \item Make sure all notation is consistent
% \item Decide if any detail should be moved to SI $\surd$
% \item Can we simplify $\boldsymbol{\nabla} \xi^\pm dS_z$ $\surd$
% \item Check $\frac{d}{dt}$ vs. $\frac{\partial}{\partial t}$  
% \item Discussion terminology, e.g. Pressure $\Pi$ (kinetic or total term) and stress $\sigma = -\Pi^c$ (just configurational part, -vs of pressure) $\surd$
% \end{enumerate}

We start with a high level description of the process in this section.
The control volume formula for surface pressure are derived by starting from the definitions of \citet{Irving_Kirkwood}.
These link the continuum expressions for density, momentum and energy to their molecular equivalents at any point in space.
The process is then a formal integral of these definitions over a volume in space, followed by the evaluation of their time evolution to get the mass and energy conservation as well as the momentum balance equations.

The volume integral follows the interface, with functional form obtained from the intrinsic surface method by fitting to the outer molecules at the interface.
For the reader only interested in applying these equations, the key formula to get pressure are given in section \ref{sec:summary} as \eq{y_equation_implementation} and \eq{z_equation_implementation}.
These have the simple interpretation of a grid which deforms and moves following the liquid-vapour interface.
We obtain all intersections of molecular interactions or particle trajectories with surfaces of the cells in that grid.
The interactions and forces over a surface are in the form of a force over area, $\boldsymbol{F}/A$, the surface pressure.
When summed over all surfaces of any enclosed volume they exactly define the change inside, a requirement of the validity of the conservation laws expressed in control volume form.
A piecewise bilinear approximation of the moving interface is used to get surface crossings more efficiently, and mapping applied to simplify calculations.
The contribution due to the movement of the interface itself is obtained by considering particles before and after movement.

\subsection{The Irving Kirkwood Equations}

The density at a given point in space can be obtained using Irving-Kirkwood's procedure \citep{Irving_Kirkwood}, defined here without the ensemble average \citep{Evans2008aa}, to give the instantaneous quantity obtained at any time in a molecular dynamics simulation,
\begin{align}
 \rho (\boldsymbol{r},t) = \displaystyle\sum_{i=1}^N m_i \delta \left( \boldsymbol{r} - \boldsymbol{r}_i  \right).
\label{densityIK}
\end{align}
Here $\rho$ is the continuum density at point $\boldsymbol{r}$ in three dimensional space and time $t$. 
The sum on the right adds the mass $m_i$ over all $N$ molecules in the system, with the Dirac delta function $\delta$ only non-zero when $\boldsymbol{r}$ is equal to $\boldsymbol{r}_i$, that is molecule $i$ is located at point $\boldsymbol{r}$.
We can define momentum, 
\begin{align}
 \rho (\boldsymbol{r},t) \boldsymbol{u} (\boldsymbol{r},t) = \displaystyle\sum_{i=1}^N m_i \boldsymbol{\dot{r}}_i \delta \left( \boldsymbol{r} - \boldsymbol{r}_i  \right),
\label{momentumIK}
\end{align}
 and energy,
\begin{align}
 \rho (\boldsymbol{r},t) \mathcal{E} (\boldsymbol{r},t) = \displaystyle\sum_{i=1}^N m_i e_i \delta \left( \boldsymbol{r} - \boldsymbol{r}_i  \right),
\label{energyIK}
\end{align}
in an analogous manner, with $\rho \boldsymbol{u}$ and $\mathcal{E}$ continuum momentum and energy respectively. Energy here is $e_i = v_i ^ 2 + 1/2 \sum_{j \ne i} \phi_{ij}$ based on half the energy of the intermolecular interaction.

\subsection{Control Volume integral}

The continuum approximation represent reality as a continuous field, obtained by taking the zero limit of an infinitesimal volume at each point.
This same concept applied to a discrete system results in, an infinitely thin and infinite large function at a point, the Dirac delta (formally a generalised function).
The Dirac delta can be thought of as a useful placeholder representing the continuum assumption in a molecular system, but has limited use in practice, particularly in software implementation.
A more tractable form of the Dirac delta function, for use in a discrete system, is obtained by the integration over an arbitrary volume to get the control volume form \citep{Smith_et_al12}.
This has the advantage that a conservative set of equations can be obtained in a molecular system.
These are directly relatable to the equivalent control volume expressions in the continuum.
More importantly for this work, the control volume shape can be chosen based on the geometry of interest.
In this work, the volume is chosen to follow the intrinsic surface as it varies in time.
The density at a point, \eq{densityIK}, can be integrated over a volume as follows,
\begin{align}
    \int_V \rho (\boldsymbol{r},t) dV = \displaystyle\sum_{i=1}^N m_i \int_V \delta \left( \boldsymbol{r} - \boldsymbol{r}_i  \right) dV, 
\label{CVdensityIK}
\end{align}
similar for momentum \eq{momentumIK},
\begin{align}
  \int_V \rho (\boldsymbol{r},t) \boldsymbol{u} (\boldsymbol{r},t) dV = \displaystyle\sum_{i=1}^N m_i \boldsymbol{\dot{r}}_i  \int_V \delta \left( \boldsymbol{r} - \boldsymbol{r}_i  \right) dV,
\label{CVmomen tumIK}
\end{align}
and energy \eq{energyIK},
\begin{align}
\int_V \rho (\boldsymbol{r},t) \mathcal{E} (\boldsymbol{r},t) dV = \displaystyle\sum_{i=1}^N m_i e_i  \int_V \delta \left( \boldsymbol{r} - \boldsymbol{r}_i  \right) dV.
\label{CVenergyIK}
\end{align}
To evaluate the integral of Eqs.\ (\ref{densityIK} - \ref{energyIK}), only the Dirac delta function must be integrated.
The volume described by the triple integral is between four flat surfaces and two faces following the shape of the intrinsic surface $\xi$,
\begin{align}
 \int_V \delta \left( \boldsymbol{r} - \boldsymbol{r}_i  \right) dV = \int_{x^-}^{x^+} \!\! \int_{y^-}^{y^+} \!\! \int_{\insurf{$-$}}^{\insurf{+}} \delta \left( x - x_i  \right) \delta \left( y - y_i  \right) \delta \left( z - z_i  \right) dz dy dx, 
\label{CV_intergal_start}
\end{align}
here the arbitrary volume is cuboidal in $x$ and $y$ directions denoted by plus and minus superscripts for top and bottom surfaces. 
The location of these surfaces are $x^+ \define x+ \Delta x/2$ and $x^- \define x - \Delta x/2$ respectively with $\Delta x$ the $x$ width of the CV with centre at point $x$, with a similar definition using $\Delta y$ as width in $y$. 
The surface in the $z$ directions is described by position $z^\pm\define z\pm \Delta z/2$ and $\xi (x, y, t)$, a continuous function of $x$, $y$ and time $t$, so surface position in $z$ denoted by $\insurf{$\pm$}(x,y,t) \define z(t)^\pm + \xi (x,y,t)$. 
In general, $\xi$ can be any function and we will make no assumption about its form until section \ref{sec:intrinsic} where a sum of trigonometric function will be used to describe the intrinsic surface between a liquid and vapour phase.

%
%\begin{align}
%\label{startingeq}
% \int_V \delta \left( \boldsymbol{r} - \boldsymbol{r}_i  \right) dV = \int_{x^-}^{x^+} \!\! \int_{y^-}^{y^+} \!\! \int_{\insurf{$-$}}^{\insurf{+}} \delta \left( x - x_i  \right) \delta \left( y - y_i  \right) \delta \left( z - z_i  \right) dz dy dx
%\nonumber \\
%=  \int_{x^-}^{x^+}  \int_{y^-}^{y^+}  m_i \delta \left( x - x_i  \right)     \delta \left( y - y_i  \right) \left[H \left( z - z_i  \right) \right]_{\insurf{$-$}}^{\insurf{+}} dy dx
%\nonumber \\
% =  \int_{x^-}^{x^+}  \int_{y^-}^{y^+}  m_i \delta \left( x - x_i  \right)     \delta \left( y - y_i  \right) \left[H \left( \insurf{+}(x,y) - z_i  \right) - H \left( \insurf{$-$}(x,y) - z_i  \right) \right] dy dx
%\nonumber \\
%=  \left[ H \left( x^+ - x_i  \right)  -  H \left( x^- - x_i  \right)\right] \;\;\;\;\;\;\;\;\;\;\;\;\;\;\;\;\;\;\;
%  \nonumber \\ 
% \times \left[ H \left( y^+ - y_i  \right)  -  H \left( y^- - y_i  \right)\right] \;\;\;\;\;\;\;\;\;\;\;\;\;\;\;\;\;\;\;
%  \nonumber \\ 
% \times \left[H \left( \insurf{+} (x_i, y_i)  - z_i \right)  -  H \left( \insurf{$-$} (x_i, y_i) - z_i  \right) \right] 
%\end{align}

As the $\insurf{$\pm$}$ limits are a function of $x$ and $y$, the $z$ integral must be evaluated first,
\begin{align}
 \int_V \delta \left( \boldsymbol{r} - \boldsymbol{r}_i  \right) dV =  \int_{x^-}^{x^+}  \int_{y^-}^{y^+}  m_i \delta \left( x - x_i  \right)     \delta \left( y - y_i  \right) 
\nonumber \\
\times \left[H \left( \insurf{+}(x,y,t) - z_i  \right) - H \left( \insurf{$-$}(x,y,t) - z_i  \right) \right] dy dx.
\end{align}
The Dirac delta is the Heaviside function upon integration, with the finite limits $\insurf{$-$}$ and $\insurf{+}$ inserted.
The next two integrals over $x$ and $y$ use the sifting property of the Dirac delta, namely $ \int \delta(x-a)f(x) = f(a)$ so the function $\xi (x,y,t)$ which describes the surface roughness becomes expressed in terms of molecular position,
\begin{align}
\vartheta_i \define  \int_V \delta \left( \boldsymbol{r} - \boldsymbol{r}_i  \right) dV =
  \left[ H \left( x^+ - x_i  \right)  -  H \left( x^- - x_i  \right)\right] \;\;\;\;\;\;\;\;\;\;\;\;\;\;\;\;\;\;\;\;\;\;\;\;\;\;\;\;\;\;\;\;\;\;\;\;\;\;\;\;\;\;\;
  \nonumber \\ 
 \times \left[ H \left( y^+ - y_i  \right)  -  H \left( y^- - y_i  \right)\right] \;\;\;\;\;\;\;\;\;\;\;\;\;\;\;\;\;\;\;\;\;\;\;\;\;\;\;\;\;\;\;\;\;\;\;\;\;\;\;\;\;\;\;
  \nonumber \\ 
 \times \left[H \left( \insurf{+} (x_i, y_i, t)  - z_i \right)  -  H \left( \insurf{$-$} (x_i, y_i, t) - z_i  \right) \right] 
= \Lambda_{xi} \Lambda_{yi} \tilde{\Lambda}_{zi},
 \label{CV}
 \end{align}
where $\vartheta_i$ a function which selects molecules inside the Heavisides, called the control volume function \citep{Smith_et_al12, Braga_et_al18}.
The $\Lambda$ notation is also introduced, a box car, or Bracewell \citep{Bracewell}, function for each direction, so e.g. $\Lambda_{xi} =  H \left( x^+ - x_i  \right)  -  H \left( x^- - x_i  \right)$ and the tilde on $\Lambda_{zi}$ indicates the function moves with the intrinsic surface $\tilde{\Lambda}_{zi} = H \left( \insurfi{+}  - z_i \right)  -  H \left( \insurfi{$-$}  - z_i  \right)$ where the subscript $i$ on $\insurf{}$ denoting the function is in terms of the molecular positions $\insurfi{$\pm$} = \insurf{$\pm$} (x_i, y_i, t)$.
As each $\Lambda$ is one if the particle is between the limits in that coordinate direction, the product $\Lambda_{xi} \Lambda_{yi} \tilde{\Lambda}_{zi}$ is therefore one if the particle is inside the volume; located between two intrinsic surface functions in the $z$ direction and bounded by two planes in the $x$ and $y$ directions. 
Any molecules outside of this volume will return a value of zero.
This control volume described by $\vartheta_i$ has a constant width and height of $\Delta x$, $\Delta y$ respectively with the depth $\Delta z$ always constant at any $x,y$ location, as the same intrinsic interface function $\xi$ is used for top and bottom surfaces, so $\insurfi{+} - \insurfi{$-$} = \Delta z$ .

\subsection{Mass}
The expression linking the control volume form of density in a continuum and molecular system of \eq{CVdensityIK} can therefore be written concisely using \eq{CV} as,
\begin{align}
 \int_V \rho (\boldsymbol{r},t) dV =    \displaystyle\sum_{i=1}^N m_i \vartheta_i.
 \label{xyint_b4simpl}
\end{align}
That is, the mass of any molecule in the enclosed region between two intrinsic surfaces in $z$ and four planes in $x$ and $y$ is contributing to the density in that control volume.

% The surface described by \eq{xyint_b4simpl} when $\zeta^+ (x_i, y_i) = sin(4\pi y_i/\Delta y) $ and $\zeta^- (x_i, y_i) = sin(2 \pi y_i/\Delta y)$ is displayed in Figure \ref{plotvolume}.
% For the geometry of Figure \ref{volume}, the expression $H \left( y^+ - y_i  \right) - H \left( y^- - y_i  \right) = 1 \forall \; \{ y_i, y \}$ and \eq{xyint_b4simpl} simplifies to,
% \begin{align}
%  \int_V \rho (\boldsymbol{r},t) dV =    \displaystyle\sum_{i=1}^N m_i \left[H \left( x^+ + \zeta^+ (y_i) - x_i  \right)
%  - H \left( x^- + \zeta^- (y_i) - x_i  \right) \right] .
% \end{align}
% \begin{figure}
%   \centering
% %    \includegraphics[width=0.5\textwidth]{./figures/CV_plot}
%       \caption{Results of a simple plot of the CV.}
%       \label{plotvolume}
% \end{figure}

We now use the control volume function to derive expressions for the fluxes over the surface of a volume, the so-called flux forms of the equations of motion.
In the continuum, the control volume analysis involves taking the time derivative of the mass in a volume to obtain the flux over the surfaces of that volume \citep{Potter_Wiggert},
\begin{align}
 \frac{d}{dt}\int_V \rho (\boldsymbol{r},t) dV =   - \oint_S \rho \boldsymbol{u} \cdot d \textbf{S} = \int_{S^+} \rho \boldsymbol{u} \cdot d \textbf{S}^+ - \int_{S^-} \rho \boldsymbol{u} \cdot d \textbf{S}^-,
 \label{continuum_mass}
\end{align}
where the $d\textbf{S} = \textbf{n} dS$ expresses an infinitesimal surface element $dS$ with surface normal $\textbf{n}$ and $\oint$ indicates the integral is over all surfaces of the volume, which here represents six piecewise surface integrals.
The molecular control volume has been shown previously derived for a uniform cuboid in space \citep{Smith_et_al12}, a sphere \citep{Heyes_et_al14}, and is extended here for a general volume between intrinsic surfaces.
Taking the time derivative of \eq{xyint_b4simpl},
% \begin{align}
%  \frac{d}{dt}\int_V \rho (\boldsymbol{r},t) dV =    \frac{d}{dt}\displaystyle\sum_{i=1}^N m_i \vartheta_i.
% \end{align}
% Taking the derivative in time, molecules mass $m_i$ does not change so only the CV functional is differentiated,
\begin{align}
 \frac{d}{dt}\int_V \rho dV =  \frac{d}{dt}\displaystyle\sum_{i=1}^N m_i \vartheta_i 
=  \displaystyle\sum_{i=1}^N m_i \frac{d \vartheta_i}{dt}  
=  \frac{d \Lambda_{xi}}{d t} \Lambda_{yi} \tilde{\Lambda}_{zi} 
+    \Lambda_{xi}  \frac{d \Lambda_{yi}}{d t}  \tilde{\Lambda}_{zi}
+  \Lambda_{xi} \Lambda_{yi} \frac{d  \tilde{\Lambda}_{zi} }{d t}.
%= \displaystyle\sum_{i=1}^N m_i \frac{d}{dt} \bigg( 
%\left[ H \left( x^+ - x_i(t)  \right)  -  H \left( x^- - x_i(t)  \right)\right] \;\;\;\;\;\;\;\; \;\;
%  \nonumber \\ 
% \times \left[ H \left( y^+ - y_i(t)  \right)  -  H \left( y^- - y_i(t)  \right)\right]   \;\;\;\;\;\;\;\; \;\;\;
% \nonumber \\ 
%\times  \left[ H \left( \insurfi{+}(t)  - z_i(t)   \right) -  H \left( \insurfi{$-$}(t)  - z_i(t)   \right) \right]  \bigg).
 \label{timeevo_mass}
\end{align}
%where .
The derivative of each of the three $\Lambda$ functions will generate two terms for the top and bottom surfaces corresponding to the six surfaces of the volume.
For example in $x$ the derivative $d \Lambda_{xi} / d t = d H(x^+ - x_i) / d t  -  d H(x^- - x_i) / d t $ can be seen to give two terms for the top and bottom surface in the $x$ direction.
Consider just the top, or $+$,  surface in $x$,
\begin{align}
\frac{d}{dt} H \left( x^+ -  x_i   \right)   = -\dot{x}_i  \delta \left(  x^+ - x_i   \right),
\label{dsurfmass_x}
 \end{align}
can be seen to be the particle's $x$ velocity $\dot{x}_i$ localised by a delta function to the $x^+$ surface, the mass flux of a particle over the surface. 
The same process can be applied for $x^-$ and $y^\pm$ surfaces.
We obtain the top surface in $z$ from $d \tilde{\Lambda}_{zi} / d t$ where both the molecular positions and the intrinsic surface $ \insurfi{$\pm$}$ depend on time,
\begin{align}
\frac{d}{dt} H \left(  \insurfi{+} - z_i   \right) = \left[ \frac{d \insurfi{+}}{dt} -  \frac{d z_i}{dt} \right] \delta \left(  \insurfi{+} - z_i   \right)
\nonumber \\
 = \left[ 
\dot{x}_i \frac{\partial  \insurfi{+}}{\partial x_i} + \dot{y}_i \frac{\partial  \insurfi{+}}{\partial y_i} + \frac{\partial  \insurfi{+}}{\partial t}
 -  \dot{z}_i \right] \delta \left(  \insurfi{+} - z_i   \right).
\label{dsurfmass_z}
 \end{align}
The time derivative of $\insurfi{+}(t) = z^+(t) + \xi (x_i(t),y_i(t), t)$ depends on particle positions $x_i$ and $y_i$ which are themselves a function of time, as well as the explicit surface time dependence.
Each of the terms have a physical interpretation, the Dirac delta function is only non zero when molecules are crossing the surface, counting at the point of crossing 
$1)$ the $z$ velocity components of the molecule $\dot{z}_i$,
$2)$ the surface curvature $\dot{x}_i \partial \xi^+ /\partial x_i$ and $\dot{y}_i \partial \zeta^+ /\partial y_i$ times the $x$ and $y$ particle's velocity at the location of a surface crossing and 
$3)$ the crossings due to surface time evolution itself $\partial \xi^+/\partial t$. 

  % and definining $w_i \equiv \dot{z}_i$ with,
% \begin{align}
% \tilde{w}^\pm \equiv \dot{z}^\pm  + \dot{x}_i \partial \zeta^\pm/\partial x_i + \dot{y}_i \partial \zeta^\pm/\partial y_i +  \partial \xi^\pm / \partial t 
% \end{align}
We evaluate \eq{timeevo_mass} using the derivatives as shown in \eq{dsurfmass_x} and \eq{dsurfmass_z} to get the time evolution of density in a molecular control volume between intrinsic surfaces,
  \begin{align}
 \frac{d}{dt}\displaystyle\sum_{i=1}^N m_i \vartheta_i  
& =- \displaystyle\sum_{i=1}^N m_i \Bigg(  \dot{x}_i \left[ \delta \left(  x^+ - x_i   \right) -    \delta \left(  x^- - x_i   \right)\right] \Lambda_{yi} \tilde{\Lambda}_{zi}  \nonumber \\
&  \;\;\;\;\;\;\;\;\;\;\;\;\;\;\;\;\;+   \dot{y}_i \left[ \delta \left(  y^+ - y_i   \right) -    \delta \left( y^- - y_i   \right)\right]  \Lambda_{xi}  \tilde{\Lambda}_{zi} 
\nonumber \\
& \;\;\;\;\;\;\;\;\;\;\;\;\;\;\;\; +  \Bigg[ \left( \dot{z}_i - \frac{d \insurfi{+}}{dt} \right) 
\delta \left(   \insurfi{+} -z_i   \right) 
-  \left( \dot{z}_i - \frac{d \insurfi{+}}{dt} \right)  \delta \left(  \insurfi{$-$} - z_i   \right)
 \Bigg] \Lambda_{xi}  \tilde{\Lambda}_{zi}   \Bigg).
 \label{dmdt_full}
 \end{align}
To simplify this expression, we introduce notation for the Dirac delta terms analogous to the continuum surface element $dS$ used in the integral,
\begin{align}
dS_{xi}^\pm \equiv  \delta \left( x^\pm  - x_i   \right) S_{xi}; \;\;\;\; dS_{yi}^\pm \equiv  \delta \left( y^\pm  - y_i   \right) S_{yi}; \;\;\;\; dS_{zi}^\pm \equiv  \delta \left( \insurfi{$\pm$}  - z_i   \right) S_{zi}.
\label{dS_z_def}
 \end{align}
with $S_{\alpha i}$ is the product of boxcar functions in the other two directions,
 \begin{align}
S_{x i} \define \Lambda_{yi} \tilde{\Lambda}_{zi}; \;\;\;\;\; S_{y i} \define \Lambda_{xi} \tilde{\Lambda}_{zi}; \;\;\;\;\; S_{z i} \define \Lambda_{xi} {\Lambda}_{yi}; \;\;\;\;\;
 \end{align}
which can be seen to define an area in space between the surfaces, e.g. $S_{x i}$ is only one if a particle is between $y^+$ and $y^-$ in $y$ and $\insurfi{+}$ and $\insurfi{+}$.
Using the definitions of \eq{dS_z_def} in the time evolution \eq{dmdt_full},
  \begin{align}
 \frac{d}{dt}\displaystyle\sum_{i=1}^N m_i \vartheta_i  
& = \displaystyle\sum_{i=1}^N  m_i   \Bigg ( \overbrace{\frac{\partial  \insurfi{+}}{\partial t} dS_{zi}^+  - \frac{\partial  \insurfi{$-$}}{\partial t} dS_{zi}^-}^{\textit{Surface Evolution}} 
\nonumber \\
&- \!\!\!\! \displaystyle\sum_{\beta \in \{ x,y,z\}} \!\!\!\! \dot{\beta}_{i} \Bigg[  dS_{\beta i}^+ -  dS_{\beta i}^- 
- \underbrace{ \frac{\partial  \insurfi{+} }{\partial \beta_i} dS_{z i}^+  + \frac{\partial  \insurfi{$-$} }{\partial \beta_i}  dS_{z i}^- }_{\textit{Curvature}} \Bigg]  \Bigg) =  - \oint_S \rho \boldsymbol{u} \cdot d \textbf{S},
 \label{mass_CV_form}
 \end{align}
where $\partial \insurfi{$\pm$}  / \partial z_i = 0$ is used to write concisely and the last equality, linking continuum and molecuale expressions, follows from \eq{continuum_mass}.
The sum over all surfaces in \eq{mass_CV_form}, allows the continuum and molecular expressions to be compared surface by surface, for $\beta=x$ and taking just the top $+$ surface this is,
  \begin{align}
 \int_{S_x^+} \rho \boldsymbol{u} \cdot d \textbf{S}_x^+ =  \displaystyle\sum_{i=1}^N  m_i \dot{x}_{i} dS_{xi}^+,
 \label{mass_CV_xsurface}
 \end{align}
and for $\beta=z$, against considering just the top surface,
  \begin{align}
 \int_{S_z^+} \rho \boldsymbol{u} \cdot d \textbf{S}_z^+ = \displaystyle\sum_{i=1}^N  m_i \Big [ \underbrace{-\dot{x}_i \frac{\partial  \insurfi{+} }{\partial x_i} - \dot{y}_i \frac{\partial  \insurfi{+}  }{\partial y_i}}_{\textit{Curvature}} + \dot{z}_{i} - \underbrace{   \!\!\!\!\!\!\!\!\!\! \!\!\!\!\!\frac{\partial  \insurfi{+}}{\partial t} \vphantom{\frac{\partial  \insurfi{+}}{\partial y_i}}   \!\!\!\!\! \!\!\!\!\!\!\!\!\!\!}_{\textit{\!\!\!\!\! \!\!\!\!\!\!\!\!\!\! Surface Evolution \!\!\!\!\! \!\!\!\!\!\!\!\!\!\!}}  \Big] dS_{zi}^+.
 \label{mass_CV_zsurface}
 \end{align}
So, the $z^+$ surface flux of mass is made up of direct fluxes, surface curvature and surface evolution components which must all be evaluated. The shorthand notation for flux over each surface is introduced,
  \begin{align}
 \frac{d}{dt}\displaystyle\sum_{i=1}^N m_i \vartheta_i  
& = -\displaystyle\sum_{i=1}^N  m_i \left[  \dot{\boldsymbol{\beta}}_{i}^+ \cdot 
  d\textbf{S}_{ i}^+ -  \dot{\boldsymbol{\beta}}_{i}^- \cdot  d\textbf{S}_{i}^-  \right] 
= -\displaystyle\sum_{i=1}^N  m_i \dot{\boldsymbol{\beta}}_{i} \cdot   d\textbf{S}_{ i},
\label{mass_CV_conservation}
 \end{align}
where
\begin{align}
\dot{\boldsymbol{\beta}}_{i}^{\pm} = 
\begin{bmatrix}
\dot{x}_i     \\
\dot{y}_i      \\
\dot{z}_i - \frac{\partial  \insurfi{$\pm$} }{\partial x_i} - \frac{\partial  \insurfi{$\pm$} }{\partial y_i} -  \frac{\partial  \insurfi{$\pm$}}{\partial t}    
\end{bmatrix} 
\textrm{  and  }
 d\textbf{S}_{i}^\pm = 
\begin{bmatrix}
 d{S}_{xi}^\pm    \\
 d{S}_{yi}^\pm    \\
 d{S}_{zi}^\pm  
\end{bmatrix} 
\end{align}
and the $\pm$ superscript is omitted in \eq{mass_CV_conservation} when expressing the difference $d\textbf{S}_{i} = d\textbf{S}_{i}^+ - d\textbf{S}_{i}^-$ or more generally could be  a shorthand to denote flux over an arbitary surface.
%  \begin{align}
% \frac{d}{dt}\displaystyle\sum_{i=1}^N m_i \vartheta_i  
% = \displaystyle\sum_{i=1}^N  m_i \bigg( \dot{x}_i \left[  dS_{xi}^+ - dS_{xi}^- \right] \; + \;  \dot{y}_i \left[ dS_{yi}^+ -  dS_{yi}^- \right] \;\;\;\;\;\;\;\;\;\;\;\;\;\;\;\;\;\;\;\;\;\;\;\;\;\;\;\;\;\;\;\;\;
% \nonumber \\
% + \;  \left[ \dot{z}_i + \dinsurfi{+} \right] dS_{zi}^+  - \left[   \dot{z}_i - \dinsurfi{-} \right]  dS_{zi}^-   \bigg)
% \label{mass_CV_form}
% \end{align}
% where the final equality introduces a vector form of the surface functional $d\textbf{S}_{i\zeta} = d{S}_{\alpha i\zeta} \; \forall \; \alpha =\{ x,y,z \}$.
%  $v_i \equiv \dot{y}_i$ and $u^+ \equiv \dot{x}^+$, \eq{timeevo_mass} becomes
% \begin{align}
%  \frac{d}{dt}\int_V \rho (\boldsymbol{r},t) dV =    \displaystyle\sum_{i=1}^N m_i \left[ \left( u^+ + v_i \frac{\partial \zeta^+}{\partial y_i} -  u_i \right)dS_{xi}^+ -  \left( u^- + v_i \frac{\partial \zeta^-}{\partial y_i} -  u_i \right)dS_{xi}^- \right]
% \end{align}
%This is in the same form as the continuum surface flux, describing the sum of momentum flux over each of the surfaces.
\subsection{Momentum}

The same integration process used for density can be applied to get the momentum in a control volume, written in integrated form as,
\begin{align}
 \int_V \rho (\boldsymbol{r},t) \boldsymbol{u} (\boldsymbol{r},t) dV =    \displaystyle\sum_{i=1}^N m_i \dot{\boldsymbol{r}}_i \vartheta_i.
\label{mom_MD_continuum_eq}
\end{align}
The equivalent time evolution of momentum for a continuum control volume gives the surface flux of momentum (convection) $\rho \boldsymbol{u} \boldsymbol{u}$ and pressure tensor denoted by $\boldsymbol{\Pi}$,
\begin{align}
\frac{d}{dt} \int_V \rho \boldsymbol{u} dV = - \oint_s \left[ \rho \boldsymbol{u} \boldsymbol{u} + \boldsymbol{\Pi} \right] \cdot d\textbf{S} = - \oint_s \left[ \rho \boldsymbol{u} \boldsymbol{u} + \boldsymbol{\Pi}^k + \boldsymbol{\Pi}^c  \right] \cdot d\textbf{S},
\label{continuum_momentum}
\end{align}
where the total pressure tensor can be split into kinetic pressure, $\boldsymbol{\Pi}^k$, and configurational pressure, $\boldsymbol{\Pi}^c$, contributions.
It is perhaps more natural to talk about the compressed state of molecules in the configurational contributions as stress, but as this is simply the negative of pressure, the term pressure is used here for both kinetic and configurational contributions.
We could also explicitly include the term for surface tension $\oint \gamma d\boldsymbol{\ell}$ in these continuum equations, but it is more convenient to include all terms in the pressure tensor and calculate the surface tension from the stress tensors using the \citet{Kirkwood_Buff} approach.

As with the mass equation, we evaluate the time evolution of \eq{mom_MD_continuum_eq} to obtain expressions for the molecular surface flux and stresses. obtained from manipulation of the Dirac delta functions,
\begin{align}
\frac{d}{dt} \int_V \rho \boldsymbol{u} dV = \frac{d}{dt} \displaystyle\sum_{i=1}^N m_i \dot{\boldsymbol{r}}_i \vartheta_i = \underbrace{\displaystyle\sum_{i=1}^N m_i \dot{\boldsymbol{r}}_i \frac{d \vartheta_i}{dt}}_{\textrm{Kinetic} } + \underbrace{\displaystyle\sum_{i=1}^N m_i \ddot{\boldsymbol{r}}_i \vartheta_i }_{\textrm{Configurational} }.
\label{Full_time_derivative_momentum}
\end{align}
% However, for presentation purposes it is simper to integrate the well known \citet{Irving_Kirkwood} equations for stress at a point,
% \begin{align}
%  \boldsymbol{\Pi} =  \bigg[  \underbrace{\displaystyle\sum_{i=1}^N m_i \left( \dot{\boldsymbol{r}}_{i} - \boldsymbol{u} \right) \left( \dot{\boldsymbol{r}}_{i} - \boldsymbol{u} \right)  \delta \left(\boldsymbol{r} - \boldsymbol{r}_i \right)}_{\boldsymbol{\Pi}^k} + \underbrace{\frac{1}{2}\sum_{i=1}^{N}\sum_{j \not= i}^{N}
%  \boldsymbol{f}_{ij} \left[ \delta \left( \boldsymbol{r} - \boldsymbol{r}_i \right) - \delta \left( \boldsymbol{r} - \boldsymbol{r}_j \right)\right] }_{\boldsymbol{\Pi}^c} \bigg],
% \label{VApressure}
% \end{align}
The Kinetic term proceeds along the same lines as density, above, which gives,
%  \begin{align}
%\displaystyle\sum_{i=1}^N m_i \dot{\boldsymbol{r}}_i \frac{d \vartheta_i}{dt} = & \displaystyle\sum_{i=1}^N  m_i \dot{\boldsymbol{r}}_i  \Bigg ( \overbrace{\frac{\partial  \insurfi{+}}{\partial t} dS_{zi}^+  - \frac{\partial  \insurfi{$-$}}{\partial t} dS_{zi}^-}^{\textit{Surface Evolution}} 
%\nonumber \\
%& + \!\!\!\! \displaystyle\sum_{\beta \in \{ x,y,z\}} \!\!\!\! \dot{\beta}_{i} \Bigg[  dS_{\beta i}^+ -  dS_{\beta i}^- 
%+ \underbrace{ \frac{\partial  \insurfi{+} }{\partial \beta_i} dS_{z i}^+  - \frac{\partial  \insurfi{$-$} }{\partial \beta_i}  dS_{z i}^- }_{\textit{Curvature}} \Bigg]  \Bigg) = \oint_S  \left[ \rho \boldsymbol{u} \boldsymbol{u} + \boldsymbol{\Pi}^k \right] \cdot d\textbf{S} 
% \label{kinetic_momentum}
% \end{align}
  \begin{align}
\displaystyle\sum_{i=1}^N m_i \dot{\boldsymbol{r}}_i \frac{d \vartheta_i}{dt} = -\displaystyle\sum_{i=1}^N  m_i \dot{\boldsymbol{r}}_i  \dot{\boldsymbol{\beta}}_{i} \cdot   d\textbf{S}_{ i} = -\oint_S  \left[ \rho \boldsymbol{u} \boldsymbol{u} + \boldsymbol{\Pi}^k \right] \cdot d\textbf{S} .
 \label{kinetic_momentum}
 \end{align}
%where $\partial \insurfi{$\pm$}  / \partial z_i = 0$ has been used to write the curvature terms concisely. 
%The convection could be separated from the kinetic pressure using the definion of velocity, although with the moving interface this is potentially more challeging.
% \begin{align}
% \int_V \boldsymbol{\nabla} \cdot \boldsymbol{\Pi}^k dV = \boldsymbol{\nabla} \cdot  \displaystyle\sum_{i=1}^N \frac{\boldsymbol{p}_{i}\boldsymbol{p}_{i}}{m_i}   \vartheta_i = \displaystyle\sum_{i=1}^N \frac{\boldsymbol{p}_{i}\boldsymbol{p}_{i}}{m_i} \cdot d\textbf{S}_{i\zeta}
% \end{align}
% where the notation $d\textbf{S}_{i\zeta}$ is identical to the one used previous in \eq{mass_CV_form}.
% Introducing the notation $\vartheta_\alpha\{A, B\} = \left[ H \left( A - r_{\alpha i}  \right) - H \left( B - r_{\alpha i}  \right)\right]$, where $\alpha$ denotes the x, y or y components. 
% The configurational pressure, $\boldsymbol{\Pi}^c$, can be integrated in the same manner noting that the order of integration can be changed as $\lambda$ is not a function of any other variable and introducing the substitution $\boldsymbol{r}_\lambda \define \boldsymbol{r}_i - \lambda \boldsymbol{r}_{ij}$,
% \begin{align}
% \int_V  \boldsymbol{\nabla} \cdot\boldsymbol{\Pi}^c dV = \boldsymbol{\nabla} \cdot \frac{1}{2}\sum_{i=1}^{N}\sum_{j \not= i}^{N}
% \boldsymbol{r}_{ij} \boldsymbol{f}_{ij} \int_0^1 \vartheta_z\{ z^+ + \zeta^+ (x_\lambda,y_\lambda), \;  z^- + \zeta^- (x_\lambda,y_\lambda) \} \vartheta_y \{y^+, y^-\}  \vartheta_x \{x^+, x^-\} d \lambda
% \end{align}
We consider the Configurational term next,
\begin{align}
\displaystyle\sum_{i=1}^N m_i \ddot{\boldsymbol{r}}_i \vartheta_i = \displaystyle\sum_{i=1}^N \boldsymbol{F}_i \vartheta_i  = \frac{1}{2}\displaystyle\sum_{i,j}^N \boldsymbol{f}_{ij} \left[\vartheta_i - \vartheta_j \right],
\label{Force_term}
\end{align}
where the $i,j$ notation is shorthand for a double sum over all $i$ and all $j$ indices.
Equation (\ref{Force_term}) is the integral of the differences of two Dirac delta functionals, the infamous IK operator \citep{Evans_Morris}.
Using the fundamental theorem of the calculus, this can be expressed in a much more convenient form,
\begin{align}
 \vartheta_i - \vartheta_j &= \int_V \left [\delta \left( \boldsymbol{r} - \boldsymbol{r}_i \right) - \delta \left( \boldsymbol{r} - \boldsymbol{r}_j \right) \right] dV 
\nonumber \\
& = \int_V \int_0^1 \frac{\partial}{\partial \lambda} \delta \left( \boldsymbol{r} - \boldsymbol{r}_\lambda \right) d \lambda dV 
= \int_0^1 \frac{\partial}{\partial \lambda}  \int_V \delta \left( \boldsymbol{r} - \boldsymbol{r}_\lambda \right) dV d \lambda 
\nonumber \\
&  =\int_0^1 \frac{\partial \vartheta_\lambda }{\partial \lambda}  d \lambda 
= \int_0^1 \frac{\partial \boldsymbol{r}_\lambda }{\partial \lambda} \cdot \frac{\partial \vartheta_\lambda }{\partial \boldsymbol{r}_\lambda} d \lambda 
=  \int_0^1\boldsymbol{r}_{ij} \cdot \frac{\partial \vartheta_\lambda }{\partial \boldsymbol{r}_\lambda} d \lambda,
\label{derivative_lambda}
\end{align}
where $\boldsymbol{r}_\lambda = \boldsymbol{r}_{i} + \lambda \boldsymbol{r}_{ij}$, representing an integration along the line of interaction between $\boldsymbol{r}_i$ and $\boldsymbol{r}_j$.
The intrinsic surface $\xi(x,y,t)$ is independent of the dummy integral along the path between molecules, $\lambda$, so we can change the order of integration allowing a volume integral of the Delta functions which
%\begin{align}
%\int_0^1 \frac{\partial}{\partial \lambda}  \int_V \delta \left( \boldsymbol{r} - \boldsymbol{r}_\lambda \right) dV d \lambda  =
%\int_0^1 \frac{\partial \vartheta_\lambda }{\partial \lambda}  d \lambda
%\label{derivative_lambda}
%\end{align}
follows the same process as \eq{CV_intergal_start} to \eq{CV} with molecular position replaced by point on line of interaction $\boldsymbol{r}_\lambda \to \boldsymbol{r}_i$, giving,
\begin{align}
\vartheta_\lambda  \define \left[H \left(  x^+ - x_\lambda  \right) - H \left(  x^- - x_\lambda  \right) \right] \;\;\;\;\;\;\;\;\;\;\;\;\;\;\;\;\;\;\;\;\;\;\;\;\;\;\;\;\;\;\;\;\;\;\;\;\;\;\;\;\;\;\,\;\;\;\,\;\; \\ \nonumber 
 \times \left[H \left( y^+ - y_\lambda  \right) - H \left( y^- - y_\lambda  \right) \right] \;\;\;\;\;\;\;\;\;\;\;\;\;\;\;\;\;\;\;\;\;\;\;\;\;\;\;\;\;\;\;\;\;\;\;\;\;\;\;\;\;\;\,\;\;\;\,\;\; \\ \nonumber 
 \times \left[H \left( \insurf{+} \left( x_\lambda, y_\lambda, t \right) - z_\lambda  \right) - H \left( \insurf{$-$} \left( x_\lambda, y_\lambda, t \right) - z_\lambda  \right) \right] = \Lambda_{x_\lambda} \Lambda_{y_\lambda} \tilde{\Lambda}_{z_\lambda}.
\end{align}
Here the control volume function selects molecular interactions through Heavisides to obtain the length of line in a control volume, by integrating along that line where any point is one when $\boldsymbol{r}_\lambda$ is inside and zero otherwise.
We denote the intrinsic interface function in terms of $\boldsymbol{r}_\lambda$ as $\insurfl{$\pm$} \equiv z^\pm + \xi (x_\lambda,y_\lambda, t)$.
The functions $\Lambda_{x_\lambda}$, $\Lambda_{y_\lambda}$ and $\tilde{\Lambda}_{z_\lambda}$ are analogous to the previous definitions with $\boldsymbol{r}_i$ replaced by $\boldsymbol{r}_\lambda$.
%The explicit time dependence is omitted here to keep notation concise.

The expression $\boldsymbol{r}_{ij} \cdot \partial \vartheta_\lambda /\partial \boldsymbol{r}_\lambda$ in \eq{derivative_lambda} is a sum over all three directions, considering the $z$ component first,
\begin{align}
{z}_{ij} \frac{\partial \vartheta_\lambda }{\partial z_\lambda} =  -{z}_{ij} \left[ dS_{z\lambda}^+ - dS_{z\lambda}^- \right] ,
\end{align}
where the $dS_{\alpha\lambda}^\pm$ term is defined following the convention used for \eq{dS_z_def} to be,
\begin{align}
dS_{x \lambda}^\pm \define \delta \left(  x^\pm  - x_\lambda  \right) S_{x\lambda}; \;\;\; dS_{y \lambda}^\pm \define \delta \left(  y^\pm  - y_\lambda  \right) S_{y\lambda}; \;\;\; dS_{z \lambda}^\pm \define \delta \left(  \insurfl{$\pm$}  - z_\lambda  \right) S_{z\lambda},
\end{align}
with $S_{x\lambda} = \Lambda_{y_\lambda}\tilde{\Lambda}_{z_\lambda}$, $S_{y\lambda} = \Lambda_{x_\lambda}\tilde{\Lambda}_{z_\lambda}$ and $S_{z\lambda} = \Lambda_{x_\lambda}\Lambda_{y_\lambda}$.
% \begin{align}
%   S_{z\lambda} = \left[H \left( x^+  - z_\lambda  \right) - H \left( x^-  - z_\lambda  \right) \right] 
%  \\ \nonumber 
% \times \left[H \left( y^+ - y_\lambda  \right) - H \left( y^- - y_\lambda  \right) \right]  
% \end{align}
The derivatives of the ${x}$ and $y$ components are slightly more complicated due to the $x_\lambda$ and $y_\lambda$ dependency in $\zeta$, so for $x$ we have,
%\begin{align}
%{x}_{ij} \frac{\partial \vartheta_\lambda }{\partial x_\lambda} =  {x}_{ij} \left[ dS_{x\lambda}^+ - dS_{x\lambda}^-  + \frac{\partial \insurfl{+} }{\partial x_\lambda} dS_{z\lambda}^+ - \frac{\partial \insurfl{-} }{\partial x_\lambda} dS_{z\lambda}^- \right] 
%\end{align}
\begin{align}
{x}_{ij} \frac{\partial \vartheta_\lambda }{\partial x_\lambda} =  -{x}_{ij} \left[ dS_{x\lambda}^+ - dS_{x\lambda}^-  - \frac{\partial \insurfl{+} }{\partial x_\lambda} dS_{z\lambda}^+ + \frac{\partial  \insurfl{-} }{\partial x_\lambda} dS_{z\lambda}^- \right].
\end{align}
%where product rule is required to address the $x_\lambda$ dependency in the $z$ surface Heaviside function 
A similar result can also be obtained for $y$.
The derivative in $x$ gives the same surface curvature terms $\partial \xi/ \partial x$ seen previously for molecular flux but in this case due to intermolecular interactions.
Combining the six surfaces of the control volume, the configurational term of \eq{Force_term} can be written as,
 \begin{align}
 \frac{1}{2}\displaystyle\sum_{i,j}^N \boldsymbol{f}_{ij} \left[\vartheta_i - \vartheta_j \right] 
%= &\frac{1}{2}\sum_{i,j}^{N} \boldsymbol{f}_{ij}  
%\int_0^1  \displaystyle\sum_{\beta \in \{ x,y,z \}} \beta_{ij} 
% \left( dS_{\beta \lambda}^+ - dS_{\beta \lambda}^- +\frac{\partial \insurfl{+} }{\partial \beta_\lambda} dS_{z \lambda}^+  - \frac{\partial \insurfl{-}  }{\partial \beta_\lambda}  dS_{z \lambda}^-  \right) d\lambda 
% \nonumber \\
= \frac{1}{2}\sum_{i,j}^{N} \boldsymbol{f}_{ij}  
\int_0^1  \boldsymbol{\beta}_{ij}  \cdot d\textbf{S}_{\lambda} d\lambda 
 = \oint_S  \boldsymbol{\Pi}^c \cdot d\textbf{S},
  \label{momentum_CV_form}
\end{align}
which is the total configurational pressure over all surfaces. 
%Again $\partial \insurfl{$\pm$}  / \partial z_\lambda = 0$ is used to write the expression concisely
Defining an analogous $\boldsymbol{\beta}_{ij}$ and $d\textbf{S}_{\lambda}$ to the kinetic flux case,
\begin{align}
\boldsymbol{\beta}_{ij}^{\pm} = 
\begin{bmatrix}
{x}_{ij}     \\
{y}_{ij}      \\
{z}_{ij} - \frac{\partial  \insurfl{$\pm$} }{\partial x_{\lambda}} - \frac{\partial  \insurfl{$\pm$} }{\partial y_{\lambda}}  
\end{bmatrix} 
\textrm{  and  }
 d\textbf{S}_{i}^\pm = 
\begin{bmatrix}
 d{S}_{x\lambda}^\pm    \\
 d{S}_{y\lambda}^\pm    \\
 d{S}_{z\lambda}^\pm  
\end{bmatrix} 
\end{align}
We can therefore take any of the surfaces, for example the configuration pressure on the $x$ surface is,
\begin{align}
 \int_{S_x^+}  \boldsymbol{\Pi}^c \cdot dS_x^+ =  \frac{1}{2}\sum_{i,j}^{N} \boldsymbol{f}_{ij}  \int_0^1 {x}_{ij}  dS_{x\lambda}^+  d\lambda,
 \end{align}
and the $z$ surface is,
\begin{align}
 \int_{S_z^+}  \boldsymbol{\Pi}^c \cdot dS_z^+ =  \frac{1}{2}\sum_{i,j}^{N} \boldsymbol{f}_{ij}  
\int_0^1 \Big[ \underbrace{ -{x}_{ij} \frac{\partial \insurfl{+} }{\partial x_\lambda} - {y}_{ij} \frac{\partial \insurfl{+} }{\partial y_\lambda} }_{\textit{Curvature}} + {z}_{ij}  \Big]  dS_{z\lambda}^+  d\lambda.
\label{momentum_CV_form_zsurface}
 \end{align}
We are now in a position to write \eq{Full_time_derivative_momentum}, the time derivative of the control volume in terms of the kinetic \eq{kinetic_momentum} and configurational \eq{momentum_CV_form} parts,
  \begin{align}
\frac{d }{dt}\displaystyle\sum_{i=1}^N m_i \dot{\boldsymbol{r}}_i \vartheta_i 
=& -\displaystyle\sum_{i=1}^N  m_i \dot{\boldsymbol{r}}_i \dot{\boldsymbol{\beta}}_{i} \cdot   d\textbf{S}_{ i}
  -  \frac{1}{2}\sum_{i,j}^{N} \boldsymbol{f}_{ij}  \int_0^1 \boldsymbol{\beta}_{ij}  \cdot d\textbf{S}_{\lambda} d\lambda 
\nonumber \\
 = & -\oint_{S} \left[ \rho \boldsymbol{u} \boldsymbol{u} + \boldsymbol{\Pi} \right] \cdot d\textbf{S},
 \end{align}
which is the total stress over every surface of the control volume.
Each of the faces of the control volume, top $+$ or bottom $-$ can be seen to define three of the components of stress tensor.
Considering the top $x$ surface, with  $dS^+_{x i}$ and $S^+_{x \lambda}$ and assuming an average pressure on the surface area $ \int_{S_x^+} \left[ \rho \boldsymbol{u} \boldsymbol{u} + \boldsymbol{\Pi} \right] \cdot d\textbf{S}_x^+  \approx \Delta S_x\left[ \rho \boldsymbol{u} u_x + \boldsymbol{\Pi}_x \right] $, with $\boldsymbol{\Pi}_x = [\Pi_{xx}, \Pi_{xy},  \Pi_{xz}]^T$ so the pressure on the $x$ surface can be written as,
  \begin{align}
  \rho \boldsymbol{u} u_x + \boldsymbol{\Pi}_x = \frac{1}{\Delta S_x}  \;\;\; \displaystyle\sum_{i=1}^N  m_i \boldsymbol{\dot{r}}_i \dot{x}_i dS_{xi}^+ + \frac{1}{2 \Delta S_x}\sum_{i,j}^{N} \boldsymbol{f}_{ij}  
\int_0^1   {x}_{ij}  dS_{x\lambda}^+  d\lambda,
\label{x_stress}
 \end{align}
which is identical to the flat surface obtained in previous work \citep{Smith_et_al12} and is consistent with a localised method of planes stress (see section \ref{sec:Implementation}).
The pressure on the $z$ surface is,
%So collecting terms for $dS^+_{z i}$ and $S^+_{z \lambda}$ and assuming an average pressure on the surface area $ \int_{S_z^+} \left[ \rho \boldsymbol{u} \boldsymbol{u} + \boldsymbol{\Pi} \right] \cdot d\textbf{S}_z^+  \approx \Delta S_z\left[ \rho \boldsymbol{u} u_z + \boldsymbol{\Pi}_z \right] $, 
  \begin{align}
  \rho \boldsymbol{u} u_z + \boldsymbol{\Pi}_z  
& = \frac{1}{\Delta S_z}  \;\;\; \displaystyle\sum_{i=1}^N  m_i \boldsymbol{\dot{r}}_i \;\;\; \Big [\overbrace{\dot{x}_i \frac{\partial \insurfi{+}}{\partial x_i} \; + \; \dot{y}_i \frac{\partial \insurfi{+}}{\partial y_i}}^{\textit{Kinetic Curvature}} \;  + \; \dot{z}_{i}
+  \overbrace{  \frac{\partial \insurfi{+}}{\partial t} \vphantom{\frac{\partial \insurfi{+}}{\partial y_i}}}^{\textit{\!\!\!\!\!\!\!\!\!\!\!\!\!\!\!Surface Evolution\!\!\!\!\!\!\!\!\!\!\!\!\!\!\!}}  \Big] dS_{zi}^+
\nonumber  \\ 
& + \frac{1}{2 \Delta S_z}\sum_{i,j}^{N} \boldsymbol{f}_{ij}  
\int_0^1 \Big[ \underbrace{ {x}_{ij} \frac{\partial \insurfl{+} }{\partial x_\lambda} + {y}_{ij} \frac{\partial \insurfl{+} }{\partial y_\lambda} }_{\textit{Configurational Curvature}} \! + \, {z}_{ij}  \Big]  dS_{z\lambda}^+  d\lambda, \;\;\;\;\;\;\;\;\;\;\;\;\;\;\;\;\;
\label{z_stress}
 \end{align}
where the equation describes the components of pressure on a control volume surface which is following the intrinsic interface, including terms due to curvature and surface evolution.
Three connected faces form a tetrahedron which is consistent with Cauchy's original definition of the stress tensor \footnote{Half of of the control volume is a tetrahedron, so the top and bottom surfaces give a different pressure tensor. In this work, one surface is no longer flat, instead following the intrinsic surface, which means we have departed from the Cauchy definition.}, giving,
\begin{align}
\left[ \boldsymbol{\Pi}_x, \boldsymbol{\Pi}_y, \boldsymbol{\Pi}_z \right]
 =
\begin{bmatrix}
 \Pi_{xx} \; & \; \Pi_{yx} \; & \; \Pi_{zx}  \\
 \Pi_{xy} \; & \; \Pi_{yy} \; & \; \Pi_{zy}    \\
\Pi_{xz} \; & \; \Pi_{yz} \; & \; \Pi_{zz} 
\end{bmatrix} \\[-15pt]
\underbrace{ }_{x \; surf}  \; \,\underbrace{ }_{y \; surf} \;\,  \underbrace{ }_{z  \; surf} \; \nonumber
\end{align}
Section \ref{sec:Implementation} will discuss how to implement this equation.

\subsection{Energy}

For completeness, the expressions for energy are stated here as they require no additional mathematics.
%For energy of the form,
% \begin{align}
%e_i = \frac{1}{2} m_i \boldsymbol{\dot{r}}_i \cdot \boldsymbol{\dot{r}}_i + \frac{1}{2} \displaystyle\sum_{j \ne i}^N \phi_{ij}
%\end{align}
The time derivative of \eq{CVenergyIK} for energy results in a similar kinetic and configurational term to the momentum equation,
  \begin{align}
\frac{d }{dt} \displaystyle\sum_{i=1}^N e_i \vartheta_i
= & \displaystyle\sum_{i=1}^N e_i \frac{d  \vartheta_i}{dt} + \displaystyle\sum_{i=1}^N \frac{d  e_i }{dt}   \vartheta_i
=  \displaystyle\sum_{i=1}^N e_i \boldsymbol{\dot{r}}_i \cdot \frac{d \vartheta_i}{d \boldsymbol{r}} + \frac{1}{2}\sum_{i,j}^{N}  \boldsymbol{\dot{r}}_i \cdot \boldsymbol{f}_{ij} \left[\vartheta_i - \vartheta_j \right]
\nonumber \\ 
= & \displaystyle\sum_{i=1}^N   e_i  \dot{\boldsymbol{\beta}}_{i} \cdot   d\textbf{S}_{ i} +  \frac{1}{2}\sum_{i,j}^{N} \boldsymbol{\dot{r}}_i\cdot \boldsymbol{f}_{ij}   
\int_0^1  \boldsymbol{\beta}_{ij}  \cdot d\textbf{S}_{\lambda} d\lambda 
\nonumber \\
 = & \oint_{S} \left[ \rho \boldsymbol{u} \mathcal{E} + \boldsymbol{\Pi} \cdot \boldsymbol{u} +  \boldsymbol{q} \right] \cdot d\textbf{S}. &
\label{Full_time_derivative_energy}
 \end{align}
Where any surface of the control volume gives the energy equation in that coordinate direction. 
Isolating and obtaining contributions for heat flux $\boldsymbol{q}$ could proceed as outlined in previous work \citep{Smith_et_al19}, although careful consideration of the evolving surface and its contribution to heat flux may be required.
This is left for future work.

%   \begin{align}
% \int_{t_1}^{t_2}  \rho \boldsymbol{u} u_z + \boldsymbol{\Pi}_z  
%  = \frac{1}{\Delta S_z}  \displaystyle\sum_{i=1}^N  m_i \boldsymbol{\dot{r}}_i & \int_{t_1}^{t_2}  \Big [\dot{x}_i \frac{\partial \insurfi{+}}{\partial x_i} + \dot{y}_i \frac{\partial \insurfi{+}}{\partial y_i}  + \dot{z}_{i}
%  \vphantom{\frac{\partial \insurfi{+}}{\partial y_i}} \Big] dS_{zi}^+ dt + \frac{1}{\Delta S_z} \displaystyle\sum_{i=1}^N  m_i \boldsymbol{\dot{r}}_i \vartheta_t
% \nonumber  \\ 
%  + \frac{\Delta t}{2 \Delta S_z}\sum_{i,j}^{N} \boldsymbol{f}_{ij}  
% & \int_0^1 \Big[  {x}_{ij} \frac{\partial \insurfl{+} }{\partial x_\lambda} + {y}_{ij} \frac{\partial \insurfl{+} }{\partial y_\lambda} + {z}_{ij}  \Big]  dS_{z\lambda}^+  d\lambda 
%  \end{align}
			% section{Theory}

% -----------------------------------------------------------------------------
\section{Implementation}
\label{sec:Implementation}

In this section, the mathematical equations are manipulated to obtain expressions which can be coded in an MD simulation.
The similarity between the kinetic and configurational terms will be highlighted, as well as the similarity in operation required for both intrinsic and flat surfaces of the volume.
It will be shown that the problem of obtaining the surface stress is reduced to obtaining the intersection of a line and a surface, a common problem in computer graphics and ray tracing. 
From equation (\ref{z_stress}), it is apparent the form of both $Kinetic\; Curvature$ and $Configurational\; Curvature$ terms are similar, with the difference between them the surface time evolution $\partial \xi^+ /\partial t$, which we consider first.

\subsection{Time Evolving Interface}
This term, $\partial \xi^+ /\partial t$, describes the change in mass, momentum or energy in a control volume as new molecules are absorbed or left behind when the intrinsic surface moves.
This can be thought of as a 2D function sweeping through space and crossing the position of the particles.
To evaluate this term, we first consider the time integration process applied in an MD simulation, i) particle positions at time $t_1$ are used for the force calculation, ii) the surface $\xi(t_1) \define \xi(x_i(t), y_i(t), t_1)$ is fixed at time $t_1$ and the evolution of the particles from $\boldsymbol{r_i}(t_1)$ to $\boldsymbol{r_i}(t_2)$ is used to get surface flux and iii) the particles are fixed at $t_2$ while the surface is evolved from $\xi(t_1)$ to $\xi(t_2)$. 
Considering the top surface in $z$, this proceeds as follows,
\begin{align}
\int_{t_1}^{t_2} \frac{\partial \insurfi{+}}{\partial t} dS_{zi}^+ d\tau = \int_{t_1}^{t_2} \frac{\partial \insurfi{+}}{\partial t} \delta \left( \insurfi{+}(t)  - z_i   \right) S_{zi} d\tau = \int_{t_1}^{t_2} \displaystyle\sum_{k=1}^{N_{roots}}  \frac{\partial \insurfi{+}}{\partial t} \frac{\delta \left( t - t_{k} \right)}{| \partial \insurfi{+}(t_k)/\partial t |} S_{zi} d\tau
\nonumber \\
%= \displaystyle\sum_{k=1}^{N_{roots}} \frac{\partial \insurfi{+}(t_k)/\partial t}{| \partial \insurfi{+}(t_k)/\partial t |} \left[H\left( t_2 - t_{k} \right) - H\left( t_1 - t_{k} \right) \right] S_{zi}(t_k) 
= \displaystyle\sum_{k=1}^{N_{roots}} sgn\left(\frac{\partial \insurfi{+}(t_k)}{\partial t} \right) \left[H\left( t_2 - t_{k} \right) - H\left( t_1 - t_{k} \right) \right] S_{zi},
%= \vartheta_t
\end{align}
using the roots of the Dirac delta function given in the Appendix \eq{DD_roots}.
The derivative of the surface can be expressed, using the definition of the partial derivative in time with $\Delta t = t_2 - t_1$,
\begin{align}
 \frac{\partial \insurfi{+}}{\partial t} = \lim_{\Delta t \to 0} \frac{\insurfi{+}(x_i(t), y_i(t), t+\Delta t) - \insurfi{+}(x_i(t), y_i(t), t)}{\Delta t} 
 \nonumber 
 \approx \frac{\insurfi{+}(t_2) - \insurfi{+}(t_1)}{t_2 - t_1},
\end{align}
so, the expression $ sgn(\partial \insurfi{+} / \partial t)  = sgn\left(\insurfi{+}(t_2) - \insurfi{+}(t_1)\right)$ as $t_2 > t_1$ and this term can be seen to simply determine the crossing direction.
Obtaining the multiple potential roots, $t_k$, of a time evolving polynomial crossing position in space is a non-trivial exercise, but the expression $sgn\left(\insurfi{+}(t_2) - \insurfi{+}(t_1)\right) \left[H\left( t_2 - t_{k} \right) - H\left( t_1 - t_{k} \right) \right]$ can be seen to be equivalent to a simple check if points $z_i(t_2)$ is crossed as the surface moves from $\insurfi{+}(t_1)$ to $\insurfi{+}(t_2)$, which can be achieved by the following functional,
\begin{align}
sgn\left(\insurfi{+}(t_2) - \insurfi{+}(t_1)\right) \left[H\left( t_2 - t_{k} \right) - H\left( t_1 - t_{k} \right) \right] S_{zi}
\nonumber \\
 = \left[H\left( \insurfi{+}(t_2) - z_i(t_2) \right) - H\left( \insurfi{+}(t_1) - z_i(t_2) \right) \right]   S_{zi} \define \vartheta_t ,
\end{align}
%which allows us to write,
%\begin{align}
% \displaystyle\sum_{k=1}^{N_{roots}} sgn\left(\partial \insurfi{+}(t_k)/\partial t\right) \left[H\left( t_2 - t_{k} \right) - H\left( t_1 - t_{k} \right) \right] S_{zi}
%\nonumber \\
%=  sgn\left(\insurfi{+}(t_2) - \insurfi{+}(t_1)\right)  \left[H\left( \insurfi{+}(t_2) - z_i(t_2) \right) - H\left( \insurfi{+}(t_1) - z_i(t_2) \right) \right]  S_{zi} \define \vartheta_t \nonumber
%%= \vartheta_t
%\end{align}
% \begin{align}
% \vartheta_t  \define sgn\left(\insurfi{+}(t_2) - \insurfi{+}(t_1)\right) S_{zi} \;\;  \;\;  \;\;  \;\; \;\;  \;\;  \;\;  \;\; 
% % \nonumber   \\ 
% % \times  \left[H \left(  x^+ - x_i  \right) - H \left(  x^- - x_i  \right) \right]  \\ \nonumber 
% %  \times \left[H \left( y^+ - y_i  \right) - H \left( y^- - y_i  \right) \right]
% \end{align}
 Note that unlike the previous control volume functionals $\vartheta_i$ and $\vartheta_\lambda$, it is possible for $\vartheta_t$ to be negative.

\subsection{Summary of Equations for Pressure}
\label{sec:summary}

In this section, the new surface pressure equations derived in the previous section are presented in a form that can be implemented in an MD simulation.
These equations will be compared to the volume average (VA) form of pressure derived in previous work \citep{Braga_et_al18}, obtained by integrating the \citet{Irving_Kirkwood} expressions over a volume following an intrinsic surface, to give,
\begin{align}
 \int_V  \big[ \rho \boldsymbol{u} \boldsymbol{u} + \PressureVA \big]  dV  =   \displaystyle\sum_{i=1}^N  m_i \dot{\boldsymbol{r}}_i  \dot{\boldsymbol{r}}_{i}  \vartheta_i +  \frac{1}{2}\displaystyle\sum_{i,j}^N \boldsymbol{f}_{ij} \boldsymbol{r}_{ij} \int_0^1 \vartheta_\lambda d \lambda.
\label{VA_relation}
\end{align}
A more general derivation of these volume average expression follows from the time evolution of \eq{Full_time_derivative_momentum}, as shown in appendix \ref{sec:surface_terms}.

 %by assuming $\partial \vartheta_i / \partial \boldsymbol{r}_i = -\partial \vartheta_i / \partial \boldsymbol{r}$ and $\partial \vartheta_i / \partial \boldsymbol{r}_\lambda = -\partial \vartheta_i / \partial \boldsymbol{r} $ to express in terms of the derivative with respect to position $\boldsymbol{r}$. 
%As shown in the appendix \ref{sec:surface_terms}, this form of the volume average omit surface curvature dependant terms.
%\displaystyle\sum_{i=1}^N m_i \dot{\boldsymbol{r}}_i \frac{d \vartheta_i}{dt} = \displaystyle\sum_{i=1}^N m_i \dot{\boldsymbol{r}}_i \frac{\partial \vartheta_i}{\partial t} + \frac{\partial \vartheta_i}{\partial \boldsymbol{r}_i} 

\begin{figure}
  \centering
    \includegraphics[width=0.8\textwidth]{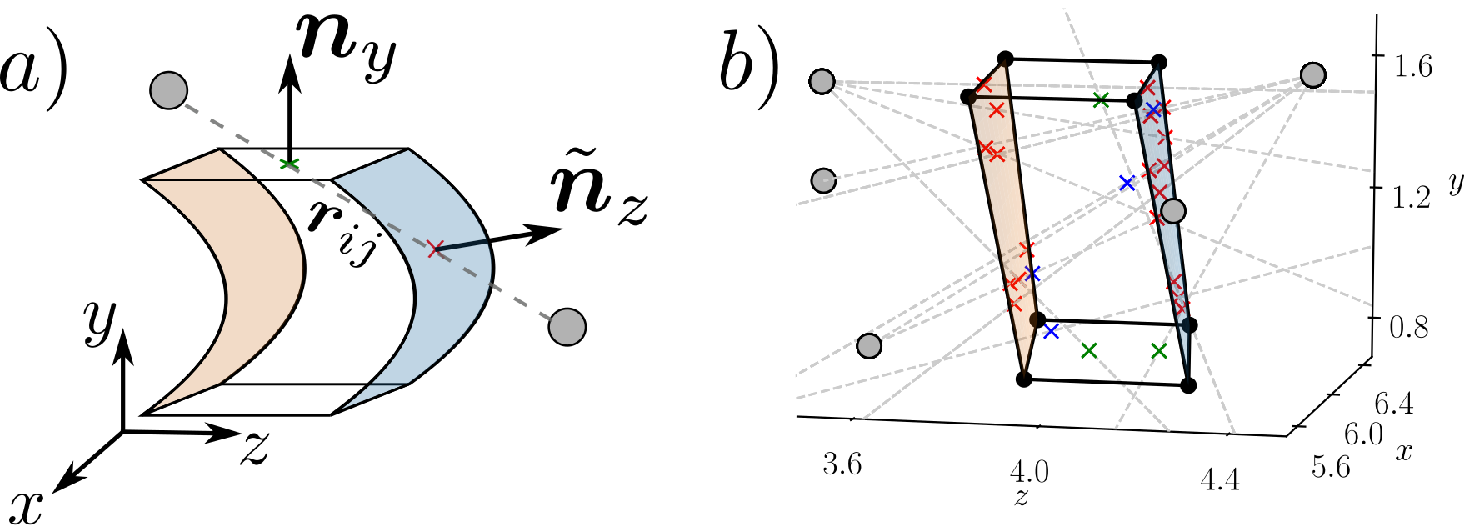}
      \caption{$a)$ A schematic showing surface normals $\boldsymbol{n}_y$ and $\tilde{\boldsymbol{n}}_z$ on an intermolecular interaction line $\boldsymbol{r}_{ij}$ contributing to pressure on both the flat surfaces via \eq{y_equation_implementation} and the intrinsic surface with \eq{z_equation_implementation} and $b)$ a random snapshot of an arbitrary control volume inside the liquid phase of an MD simulation, showing all the contributions used in the calculation of configurational pressure; with blue crosses on the $x$ surface, green crosses on the $y$ and red crosses for a bilinear patch of the intrinsic surface with the top $\xi^+$ surface coloured in light blue and the bottom $\xi^-$ coloured light red.}
      \label{normal_and_intersection_schematic}
\end{figure}

The equations for pressure on a flat surface \eq{x_stress} and an intrinsic surface \eq{z_stress} must be integrated so they can be used in a molecular simulation.
The process of taking this integral is given in appendix \ref{sec:Appendix_CV}, with just the final forms stated here.
The equation for pressure on the flat control volume faces, here the $y^+$ surface is chosen, is shown in the appendix \ref{sec:Appendix_CV} to be,
  \begin{align}
\int_{t_1}^{t_2}  \left[ \rho \boldsymbol{u} u_y + \StressSurf{\!}_y^k \right] dt = & \frac{1}{\Delta S_y}  \displaystyle\sum_{i=1}^N  m_i \boldsymbol{\dot{r}}_i  
 \frac{ \boldsymbol{r}_{i_{12}} \cdot \boldsymbol{n}_y }{|\boldsymbol{r}_{i_{12}} \cdot \boldsymbol{n}_y |} 
\! \left[H \left( \frac{y^+- y_{i_2}}{y_{i_{12}}} \right) - H \left( \frac{y_{i_1}-y^+}{y_{i_{12}}} \right) \right] \Lambda_x(t_k) \tilde{\Lambda}_z(t_k) 
  \nonumber \\
\StressSurf{}_y^c = & \frac{1}{2} \frac{1}{ \Delta S_y}\sum_{i,j}^{N} \boldsymbol{f}_{ij}  
\frac{\boldsymbol{{r}}_{ij} \cdot {\boldsymbol{n}}_y }{|\boldsymbol{r}_{ij} \cdot {\boldsymbol{n}}_y|} 
\; \left[H \left( \frac{y^+- y_j}{y_{ij}} \right) - H \left( \frac{y_i- y^+}{y_{ij}} \right) \right] \Lambda_x(\lambda_k) \tilde{\Lambda}_z(\lambda_k).
\label{y_equation_implementation}
 \end{align}
Equation (\ref{y_equation_implementation}) is written in this form to emphasise the expression is the molecular pressure tensor $m_i \boldsymbol{\dot{r}}_i \boldsymbol{r}_{i_{12}} $ and  $\boldsymbol{f}_{ij}  \boldsymbol{{r}}_{ij}$ dotted with the surface normal $\boldsymbol{n}_y$.
The $\boldsymbol{r}_{i_{12}} = \boldsymbol{r}_{i_{2}} - \boldsymbol{r}_{i_{1}}$ term is the vector from the position of molecule $i$ at time $t_1$ to its position at $t_2$.
This Heaviside functions $H$ in the square brackets check if the position $y_{i_1}$ before moving and $y_{i_2}$ after are on opposite sides of the surface at $y^+$.
These are equivalent to the method of planes (MOP) form of stress, as demonstrated in the Appendix \ref{sec:Appendix_CV}, but localised to a control volume surface by $\Lambda_x \tilde{\Lambda}_z$. 
Recall that the Lambda functions are boxcar or Bracewell functions checking if a point is between two points using Heaviside functions $\Lambda_x(a) = H(x^+ - a) - H(x^- - a)$ and $\tilde{\Lambda}_z(a) = H(\insurfi{+} - a) - H(\insurfi{$-$} - a)$.
This control volume surfaces and associated normals are shown in Figure \ref{normal_and_intersection_schematic} $a)$ for an intermolecular interaction crossing the $y^+$ surface by green crosses.
An example of the actual interactions from an MD simulation at a single timestep are also shown in Figure \ref{normal_and_intersection_schematic} $b)$ with blue and green crosses denoting the $x$ and $y$ surfaces, respectively.
The expressions for the roots $t_k$ and $\lambda_k$ can be obtained analytically for this flat surface case, with expression given in the Appendix \ref{sec:Appendix_CV}.
The expression for stress on the intrinsic surface $z^+$ is, 
\begin{align}
\int_{t_1}^{t_2}  \rho \boldsymbol{u} u_z + \StressSurf{}_z^k  dt
 = & \frac{1}{\Delta S_z}  \displaystyle\sum_{i=1}^N  m_i \boldsymbol{\dot{r}}_i  
 \frac{\boldsymbol{r}_{12} \cdot \tilde{\boldsymbol{n}}_z }{|\boldsymbol{r}_{12} \cdot \tilde{\boldsymbol{n}}_z |} 
 \displaystyle\sum_{k=1}^{N_{roots}}  \left[ H \left( 1- t_{k} \right) - H \left(  - t_{k} \right) \right] \Lambda_x(t_k)  \Lambda_y(t_k) 
  \nonumber \\
  + & \frac{1}{\Delta S_z} \displaystyle\sum_{i=1}^N  m_i \boldsymbol{\dot{r}}_i  \vartheta_t %sgn\left(\insurfi{+}(x_i(t_2), y_i(t_2), t_2) - \insurfi{+}(x_i(t_2), y_i(t_2), t_1)\right) \Lambda_x(t_2)  \Lambda_y(t_2) 
\nonumber  \\ 
 \StressSurf{}_z^c= & \frac{1}{2 \Delta S_z}\sum_{i,j}^{N} \boldsymbol{f}_{ij}  
\frac{\boldsymbol{{r}}_{ij} \cdot \tilde{\boldsymbol{n}}_z }{|\boldsymbol{r}_{ij} \cdot \tilde{\boldsymbol{n}}_z|} \displaystyle\sum_{k=1}^{N_{roots}}  \left[ H \left( 1- \lambda_k \right) - H \left(  - \lambda_k \right) \right] \Lambda_x(\lambda_k)  \Lambda_y(\lambda_k),
\label{z_equation_implementation}
 \end{align}
where, as in the flat surface, the expression for $m_i \boldsymbol{\dot{r}}_i \boldsymbol{r}_{i_{12}} $ and  $\boldsymbol{f}_{ij}  \boldsymbol{{r}}_{ij}$ are dotted with the surface normal. 
In this case, the normal to the intrinsic interface $ \tilde{\boldsymbol{n}}_z(x,y) = \frac{ \boldsymbol{\nabla}_\alpha  \left(   \xi - z_\alpha \right)  }{|| \boldsymbol{\nabla}_\alpha  \left(   \xi - z_\alpha \right) ||}$.
The normal $\tilde{\boldsymbol{n}}_z$ is shown on the schematic of Figure \ref{normal_and_intersection_schematic} $a)$ and the MD simulation of Figure \ref{normal_and_intersection_schematic} $b)$ with the many surface intersections shown as red crosses.
The intersection of the interaction line and the intrinsic surface is required to evaluate the expression of \eq{z_equation_implementation}, an identical roots-finding process for crossing time $t_k$ or point on intermolecular interaction $\lambda_k$.  
Given the multiple possible crossings, a closed form expression is not possible. 
An approach based on approximating the surface as a set of bilinear patches is used here which allows application of an efficient root calculation from the ray-tracing literature \citep{Ramsey_bilinear_2004}. 
The computations are then fast enough to be part of a molecular simulation.
Once the roots $t_k$ and  $\lambda_k$ have been found, they can be put into the $\Lambda$ functions to check if they are within the limits of a given control volume surface.
In practice, integer division can be used to speed this process up, assigning the crossing to a cell, as discussed in the next section.
Finally, the surface evolution term is obtain by checking if a particles positions $\boldsymbol{r}_i(t_2)$ is in a given volume before and after the surface has evolved. 

The result is a concise expression for the momentum flux over all surfaces, written as,
  \begin{align}
\overbrace{ \displaystyle\sum_{i=1}^N \left[ m_i \dot{\boldsymbol{r}}_i(t_2) \vartheta_i(t_2) -\dot{\boldsymbol{r}}_i(t_1) \vartheta_i(t_1) \right]  }^{\textit{Accumulation}}
= -\!\! \sum_{\alpha=\{x,y,z\}} \! \frac{1}{\Delta S_\alpha}  \Bigg[ \overbrace{\displaystyle\sum_{i=1}^N  m_i \boldsymbol{\dot{r}}_i  \left( {r}_{\alpha 12}  \left[dS^+_{\alpha t_k} - dS_{\alpha t_k}^- \right] + \vartheta_t  \right)  }^{\textit{Advection}}
\nonumber \\
+\underbrace{ \frac{\Delta t}{2 }\sum_{i,j}^{N} \boldsymbol{f}_{ij}  
{r}_{\alpha ij}  \left[dS^+_{\alpha \lambda_k} - dS^-_{\alpha \lambda_k} \right]  }_{\textit{Forcing}} \Bigg],
\label{Labelled_CV_terms}
 \end{align}
where both sides are integrated over time with the force term shown integrated with the midpoint rule for simplicity $\int dt \approx \Delta t$, consistent with the Verlet integration scheme used to propagate molecular positions and velocities.
The function to get crossings on a surface is denoted for time $s=t$ and space $s=\lambda$ with,
\begin{align}
dS^\pm_{\alpha s_k} =  \frac{\tilde{n}_{\alpha} }{|\boldsymbol{r}_{12} \cdot \tilde{n}_\alpha |}  \displaystyle\sum_{k=1}^{N_{roots}}  \left[ H \left( 1- s_{k} \right) - H \left(  - s_{k} \right) \right] \Lambda_\beta(s_k)  \Lambda_\gamma(s_k),
\end{align}
where the $\beta$ and $\gamma$ are the orthogonal directions to $\alpha$, so if $\alpha=x$, $\beta=y$ and $\gamma=z$, and the normal $\tilde{n}_{\alpha}$ can be for a flat or intrinsic surface with roots $s_k$ obtained for that surface.
This same expression is valid for both kinetic and configurational terms on both curved and flat surfaces, with the simple interpretation of checking if a crossing is on the surface of a control volume.

The six surface pressures summed for any arbitrary control volume consist of terms due to intermolecular forces, labelled $Forcing$ in \eq{Labelled_CV_terms} as well as kinetic molecular crossings due to both molecular motion and surface movement labelled $Advection$.
These are exactly equal, to machine precision, to the change in momentum in the volume, called $Accumulation$, demonstrated in section \ref{sec:Results and Discussion}.

We now discuss the process of how to obtain these surface crossing terms efficiently as part of an MD simulation.

\section{Method}
\label{sec:Methods}

We use molecular dynamics (MD) simulation to model the liquid vapour interface, with a shifted Lennard Jones potential,
\begin{align}
\phi(r_{ij}) = \epsilon \left[ \left(\frac{\sigma}{r_{ij}}\right)^{12} - \left(\frac{\sigma}{r_{ij}}\right)^{6} \right] - \phi(r_c); \;\;\;\;\; r_{ij} < r_c,
\end{align}
with cutoff $r_c=2.5$, which is shorter than required to give good agreement with the experimental measurements of the surface tension in argon \citep{reynolds1979thermodynamic, Bo_Shi_Thesis, Smith_et_al16} but chosen to allow more efficient simulations.
The force on particle $i$ is obtained from the sum of the gradient of this potential due to interaction with all $N$ other particles $\boldsymbol{F}_i = \sum_{j \ne i}^N\boldsymbol{f}_{ij} = -\sum_{j \ne i}^N\boldsymbol{\nabla} \phi(r_{ij})$
The simulations are run using the Flowmol MD code which has been validated extensively in previous work \citep{Smith_Thesis}
The time integration is achieved by velocity Verlet with a timestep $\Delta t = 0.005$.
\begin{align}
\boldsymbol{r}_i(t+\Delta t) & =\boldsymbol{r}_i(t) + \Delta t \boldsymbol{v}_i(t+\Delta t/2) \nonumber \\
\boldsymbol{v}_i(t + \Delta t /2 ) & = \boldsymbol{v}_i(t- \Delta t/2) + \Delta t  \boldsymbol{F}_i(t)
\end{align}
The system is initialised as a liquid-vapour coexistence by creating an FCC lattice and removing molecules until the desired density is obtained.
The system is periodic in all directions, with $Lx = Ly = 12.7$ while the surface normal direction $Lz= 47.62$.
The middle $40\%$ of the domain is designated to be initialised as liquid with a density of $\rho_l = 0.79$ and the remaining domain is set to a gas density of $\rho_g = 0.0002$. 
This results in a system with $N=2635$ molecules.
The system is then run in the NVT ensemble, controlled by a Nos\`{e} Hoover thermostat at a temperature setpoint of $T_s=0.7$ for 100,000 timesteps to equilibrate.
The run is then restarted as an NVE ensemble and run for sufficient time to collect well resolved statistics.

In order to obtain the intrinsic interface, a cluster analysis is first used to identify the connected molecules defined to be a liquid cluster, shown in Fig \ref{schematic} $a)$.
In this simulation, the liquid region tends to be located in the centre of the domain and the cluster analysis identifies all connected molecules which are within the Stillinger cutoff length $r_d=1.5$ from each other.
A linked list is built of all molecules, before any which have fewer than three neighbours are discarded as not part of the cluster, giving $N_{\ell}$ liquid particles.
The cluster is then used to fit the intrinsic surface, as detailed in section \ref{sec:intrinsic}, with the fitting performed on the surface on the right-hand side of the cluster.

\subsection{Intrinsic Surface}
\label{sec:intrinsic}

\begin{figure}
  \centering
    \includegraphics[width=1.0\textwidth]{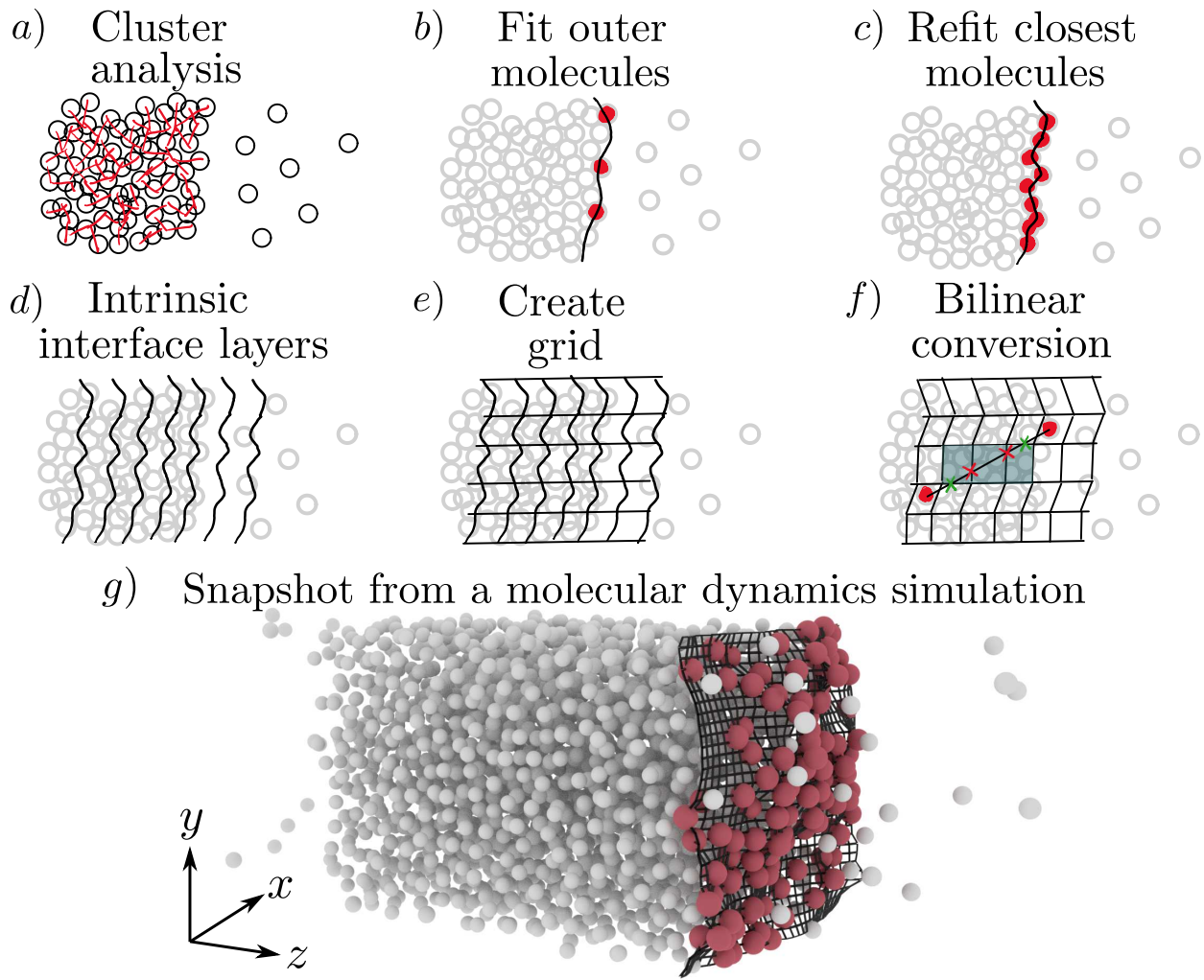}
      \caption{A schematic outlining the process of defining the surface fluxes starting by  $a)$ getting the liquid cluster, $b)$ fitting the initial surface to a number of outermost particles, with 3 show here of the $3\times 3$ taken $c)$ refining by including all closest molecules until surface density is a target value $\rho_s = n_0/A$, $d)$ defining a set of layers with uniform offset $\Delta z$ from the surface $e)$ Defining a grid by dividing the domain into a uniform grid of $\Delta x$ and $ \Delta y$ cells and $f)$ approximating the intrinsic interface as bilinear patches for each cell, with the calculation of crossings shown for flat surface as green crosses, the shaded region showing the cells between these flat surface crossings that are checked for intrinsic surface crossings with the two possible crossings shown by red crosses. A snapshot of the actual intrinsic surface ontained in the molecular dynamics simulation is shown in $g)$ with red pivot molecules and the intrinsic surface shown as a black grid split into $32 \times 32$ bilinear patches.}
      \label{schematic}
\end{figure}

Up until this section, no assumption has been made about the functional form of surface $\xi(x,y,t)$.
To give a general form, the intrinsic surface method (ISM) approximates the liquid-vapour interface using a Fourier series representation,
\begin{align}
\xi(x,y,t)= \displaystyle\sum_{\boldsymbol{k} < k_u}  \hat{\xi}_{\boldsymbol{k}}(t) 
\exp\left({2 \pi i  \boldsymbol{k} \cdot \boldsymbol{r}_{||} }\right),
\label{eqn:meth2}
\end{align}
where $\hat{\xi}_{\boldsymbol{k}}(t)$ are the amplitudes associated with each wavevector, a function of time as they are refitted to the surface every time the molecules in the system evolve,  with wave vector $\boldsymbol{k} =(\mu/Lx, \nu/Ly)$, parallel surface components $\boldsymbol{r}_{||} = (x,y)$ and the number of wavelengths calculated from the system size $k_u = nint(\sqrt{Lx Ly}/\lambda_{u})$ based on $\lambda_{u}$ the minimum wavelength of the surface, set to intermolecular spacing for a Lennard Jones fluid in this work, so $\lambda_{u}\approx\sigma$, as used in previous work \citep{Chacon2003aa}.
%The wavenumber is $k = (2 q_u + 1)  (\mu + q_u) + (\nu + q_u)$
The surface fitting functions can be made simpler by expressing in terms of just the real components,
\begin{align}
\xi(x,y,t)=\displaystyle\sum_{\mu=-k_u}^{k_u} \displaystyle\sum_{\nu=-k_u}^{k_u}  a_{\mu \nu}(t) f_\mu (x) f_\nu(y),
\label{Matrix_eqn}
\end{align}
where $f_\mu (x) = \cos(k_x x)$, $f_{-\mu} (x) = \sin(k_x x)$ and $ a_{\mu \nu}$ is the matrix of surface wavenumbers.
To fit the intrinsic surface, the square of the difference between surface molecules, called pivots, and the intrinsic surface function $\xi$ is defined to be,
\begin{align}
W = \frac{1}{2}\displaystyle\sum_{p=1}^{N_p} \left[ z_p - \xi(x_p, y_p) \right]^2 +  \psi \tilde{A},
\label{Minimise}
\end{align}
which is the function to be minimised as part of a least squares fit. 
An extra constraint $\psi$ is included to prevent overfitting by ensure the intrinsic area $\tilde{A}$ 
%\begin{align}
%\tilde{A} = A + \frac{4 \pi^2}{2} \displaystyle\sum_{\mu=-k_u}^{k_u} \displaystyle\sum_{\nu=-k_u}^{k_u}   a_{\mu \nu} [ \mu^2 + \nu ^2] 
%\end{align}
does not become too large, where $\psi=1\times10^{-8}$ in line with values in the literature \citep{Bresme2008ac}. 
This is a linear algebra problem, as discussed in Appendix \ref{sec:LA}.
The fitting of the interface is done in stages, with the first fit to a subset of molecules of the cluster obtained by defining a grid, usually $3 \times 3$ in $x$ and $y$, although a finer grid can speed up fitting \citep{Longford_et_al18}, and obtaining the outermost molecule in each gridcell.  
The fitting to these $N_p=9$ molecules is shown schematically in Fig \ref{schematic} $b)$. 
The resulting zeroth interface guess, $\xi^0$, is then interrogated at the location of each molecules with the difference sorted in order of $\xi^0(x_i, y_i)-z_i \forall N_{\ell}$. 
The closest molecules are then included and used to refit the surface, including the original $9$ and incorporating an additional one so $N_p=10$.
This can be done in batches to improve efficiency with a number of the closest molecules added at the same time, giving identical results in most cases to the process applied molecules by molecule.
With the new surface, denoted by $\xi^1$, the process is repeated and continues until a desired density of $\rho_s=N_0/A$ is reached.
The density of $\rho_s=0.7$ is shown to minimise molecular turnover in  \citet{Chacon_et_al_2009} so is used in this work.
When the number of pivots is $N_p =\rho_s A$ is reached the process is stopped, with an interface as shown in Fig \ref{schematic} $c)$. 

Once an intrinsic surface has been obtained, layers can be defined by using an offset $\Delta z$ to define binning regions.
These can be used to bin quantities such as position, velocity or fraction of intermolecular interaction to obtain density, momentum or pressure respectively at fixed distance from the surface as shown in Fig \ref{schematic} $d)$.
This work extends this approach by using localisation to a grid of cells in the surface tangential directions, as shown in  Fig \ref{schematic} $e)$. 
This allows properties as a function of position on the interface to be evaluated and could be applied to explore surface transport, local evaporation and Maragnoi effects, among other things.
In order to evaluate local quantities efficiently, the grid is replaced by a bilinear approximation shown in Fig \ref{schematic} $f)$, which simplifies the calculation of the point of intersection shown by red crosses and the process of mapping to get binned contributions. 
Section \ref{sec:bilinear} discussed this implementation in more detail.

%where $f_\mu (x) = \cos(2\pi \mu x /L_x)$ and $f_{-\mu} (x) = \sin(2\pi \mu x /L_x)$ .
% This is not right, I think it misses cross terms
%\begin{equation}
%\xi(x,y)= \displaystyle\sum_{\mu=-q_u}^{-1} \displaystyle\sum_{\nu=-q_u}^{-1} a_{\mu \nu} \ sin \left(\frac{2\pi |\mu| x }{L_x} \right)   \sin \left(\frac{2\pi |\nu| y }{L_y} \right) + \displaystyle\sum_{\mu=0}^{q_u} \displaystyle\sum_{\nu=0}^{q_u}  a_{\mu \nu}  \cos \left(\frac{2\pi \mu x }{L_x} \right)   \cos \left(\frac{2\pi \nu y }{L_y} \right),
%\end{equation}
%where $\mu$ and $\nu$ are integers

\subsection{Mapping of Points and Intersections}
\label{sec:Mapping}
\begin{figure}
  \centering
    \includegraphics[width=0.8\textwidth]{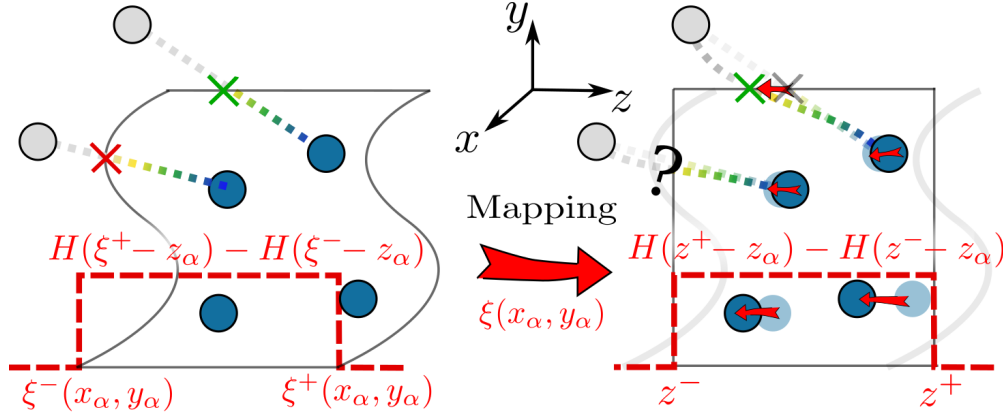}
      \caption{A schematic view of the three operations required to obtain density, momentum and pressures including i) binning molecular position $z_\alpha \to z_i$, ii) binning lines between molecules approximated as a series of points $z_\alpha \to z_\lambda$ and iii) obtaining the location of surface crossings $z_\alpha \to z_k$. The two Heaviside functions are essentially a binning operation, checking if a point is between $\xi^{-}$ and $\xi^{+}$ with $\xi^{\pm} = z^\pm + \xi$, shown in red, which can be mapped to a simpler function on the right by subtracting the intrinsic surface value at the particle locations $\xi(x_i,y_i,t)$ or point on line $\xi(x_{\lambda},y_{\lambda}, t)$. This mapping changes the calculation to a simple check if molecules are between flat surfaces $z^-$ and $z^+$ by integer division. Intersections on flat surfaces, denoted by a green cross, can be mapped in the same way to find the surface of the CV they apply, obtained by \eq{y_equation_implementation}. Mapping is not possible for crossings of the intrinsic surface, denoted by a red cross before mapping, but a question mark after to emphasise a root finding must be employed.}
      \label{CV_function}
\end{figure}

The control volume functions $\vartheta_i$ and $\vartheta_\lambda$ mathematically determine if a given point $\boldsymbol{r}_\alpha$ is located in any given cell as it moves with the intrinsic interface. 
In practice, a contiguous grid of cells is used, so instead of testing every point with every cell, the point locations are assigned to an appropriate cell using integer division.
In pseudocode for this is just,
\begin{lstlisting}
%\textcolor{function}{function}% get_cell(r)
     cell(1) = ceiling(r(1)/cellside(1))  # Function %\textcolor{comment}{$\Lambda_x$}% 
     cell(2) = ceiling(r(2)/cellside(2))  # Function %\textcolor{comment}{$\Lambda_y$}% 
     cell(3) = ceiling(r(3)/cellside(3))  # Function %\textcolor{comment}{$\Lambda_z$}% 
     return cell
%\textcolor{function}{end function}% get_cell
\end{lstlisting}
where $r(\alpha)$ is $\boldsymbol{r}_\alpha$, the point in three dimensions and rounded division by cellside$(\alpha)$ returns an integer which is the index of the cell where the point is located.
Quantities can then be added to a cell using this index, including mass, velocity, energy when the point is a molecular location or stress and work when the point is on the intermolecular line of interaction. 
This can be seen to be a computational implementation of the boxcar functions $\Lambda_x$, $\Lambda_y$ and $\Lambda_z$ defined in \eq{CV}, with no tilde in the $z$ direction as this is for a flat surface.
The intrinsic surface form $\tilde{\Lambda}_z$ can then be implemented by a simple mapping of the point followed by rounding to obtain the cell index. 
\begin{lstlisting}
%\textcolor{function}{function}% get_intrinsic_cell(r)
     cell(1) = ceiling(r(1)/cellside(1)) # Function %\textcolor{comment}{$\Lambda_x$}% 
     cell(2) = ceiling(r(2)/cellside(2)) # Function %\textcolor{comment}{$\Lambda_y$}% 
     # Function %\textcolor{comment}{$\tilde{\Lambda}_z$}% 
     cell(3) = ceiling((r(3)-xi(r(1),r(2)))/cellside(3)) 
     return cell
%\textcolor{function}{end function}% get_intrinsic_cell
\end{lstlisting}
where xi is $\xi$ at each timestep, a function which returns the $z$ position of the intrinsic surface given inputs of position $x$ and $y$.
These can be $x_i, y_i$, $x_\lambda, y_\lambda$, or $x_k, y_k$ shown in Figure \ref{CV_function} where this mapping is applied to molecular positions, intermolecular interaction modelled in a piecewise manner or surface crossings, respectively.
This has the convenient property that the intrinsic surfaces mapping can be switched on in a code by simply swapping a function pointer from the get\_cell function to the get\_intrinsic\_cell function.
For large domains in the surface tangential directions, the time taken by function xi to evaluate surface position can be prohibitive as it requires a sum over the two-dimensional Fourier surface, a calculation of order $\mathcal{O}(4 k_u^2)$ with $k_u$ scaling as $L$ (assuming $L_x=L_y=L$) so each surface evaluation is $\mathcal{O}(L^2)$.
The position of the surface at the location of every molecule is pre-calculated for efficiency to get density, momentum, temperature, potential energy and the kinetic part of the pressure tensor. 
Interactions between molecules must be obtained on the fly by discretising the line of interaction between two molecules and using the surface value at each segment of the line, as shown in Fig \ref{CV_function} by the dotted interaction lines.
Some optimisations can be introduced, for example a quick estimate of the number of bins crossed between two points is used to minimise the number of segments in the line and the intermolecular stress and work calculation are performed at the same time.
However, this still requires a larger number of evaluations of the intrinsic surface as each interaction crosses $\mathcal{O}(20)$ bins, for an average of $\mathcal{O}(100)$ interactions per particle (with $r_c=2.5$) required for each of the $N$ particles.
This is even worse for the control volume flux calculation, which requires the intersection of the intrinsic interface and the line of interaction, computational prohibitive as the intersection point is required to machine precision if we want to verify exact conservation of momentum.
Even an efficient root finding algorithm with good initial estimate requires the surface to be evaluated many times to converge for every one of the $\mathcal{O}(20)$ crossed surface by every one of the $\mathcal{O}(100)$ interactions. 
These rough numbers clearly depend on bin resolution, cut off length as well as the density of liquid and vapour states, but even in low resolution cases appear to represent a limiting step computationally.
A fast evaluation is therefore of paramount importance for intrinsic interface tracking calculation to be used as part of an MD simulation, achieved here using a bilinear approximation.

\subsection{Bilinear Approximation for Surface Crossings}
\label{sec:bilinear}
%Evaulating a point on the intrinsic surface is an $O(n_{x}\times n_{y})$ operation as the Fourier coeffiicents require a recalcution from the entire global set of global coefficients.
%
%For a small system $N=2000$ particles the number of interactions is order $100N$ with a cutoff length of $r_c=4.5$ and this is required for every timestep.
%Althugh this could potentially be an ideal case for GPU acceleration, it is prohibitve on a CPU.

The surface is replaced by a piecewise continuous bilinear surface, sampling the intrinsic surface at an arbitrary accuracy.
This uses a bilinear patch of the form,
\begin{align}
\xi^{BL} (x,y) = a_0 + a_x x + a_y y + a_{xy} x y
\end{align}
where the coefficients are obtained by solving,
\begin{align}
\begin{bmatrix}
a_0       \\
a_x       \\
a_y     \\
a_{xy}
\end{bmatrix}
=
\begin{bmatrix}
1 & x_1 & y_1 & x_1 y_1          \\
1 & x_1 & y_2 & x_1 y_2          \\
1 & x_2 & y_1 & x_2 y_1          \\
1 & x_2 & y_2 & x_2 y_2  
\end{bmatrix}^{-1}
\begin{bmatrix}
\xi(x_1,y_1)       \\
\xi(x_1,y_2)      \\
\xi(x_2,y_1)    \\
\xi(x_2,y_2)
\end{bmatrix}
\end{align}
for each patch of the intrinsic interface.
Once we have the expression for the bilinear surface, this can be used in the mapping for both particle position and lines between molecules.
In order to get surface crossings, we use the fast bilinear-patch line intersection algorithm \citep{Ramsey_bilinear_2004} which equates the solution of a line of interaction $r_s = r_1 + s r_{12}$ and the bilinear patch $\xi^{BL}$ to obtain a quadratic equation. 
Solving this provides up to two crossings for a given surface, with some care required for zero or complex roots.
As the surface is split into piecewise bilinear patches, this also has the advantage that each control volume has its own bilinear surface and the conservation check of \eq{Labelled_CV_terms} can be performed on each cell.
The derivatives can also be obtained analytically $\partial \xi^{BL}/\partial x=a_{x} + a_{xy}y$ and $\partial \xi^{BL}/\partial y = a_{y} + a_{xy}x$, although these are discontinuous at the interface of contiguous bilinear patches.
In practice this does not represent a problem for the surface pressures as the $\Lambda_\alpha$ functions ensure all quantities are per cell.
The advantage of supplementing the full intrinsic surface with the bilinear approximation, instead of directly fitting a bilinear surface, is the calculation of quantities from the full intrinsic surface $\xi$ are available analytically, if needed, with the bilinear solution used as a starting point.
Bicubic or higher order surface approximations could also be used but would not be able to take advantage of the fast interaction calculation\citep{Ramsey_bilinear_2004} for line and bilinear patch.

Determining the number of crossings of an intermolecular interaction and the 3D grid following the intrinsic surfaces (see Fig \ref{schematic} $f)$) turns out to be a non-trivial exercise.
This is because it is not possible to predict a priori how many times the intermolecular lines will cross an arbitrary intrinsic surface split into interconnected $\xi^{BL}$ patches from just the knowledge of the starting and ending points.
To solve this, for each interaction we first obtain all the crossings locations on all flat surfaces in $x$ and $y$, made possible as we have the closed form expression from \eq{y_equation_implementation}.
By ordering all intermediate flat surface ($x$ and $y$) crossings along the line of interaction, including the two particle positions at the end, we can take successive pairs and are guaranteed any intrinsic surface crossings between a pair must be on the same $\xi^{BL}$ patch. 
This is shown schematically in Fig \ref{schematic} $f)$ with green crosses denoting a pair of flat plane crossings, the shaded region shows the row of cells in between these two flat crossings to be checked for intrinsic surface intersections and the red crosses denoting two found intrinsic crossings.
As each grid volume must have at least two crossings or contain an end point (molecule position), by stepping in pairs we ensure all surface crossings have been obtained by this process.
This has the added advantage that the coefficients of the bilinear surface, $\xi^{BL}$, can be loaded once for each row, the shaded region in Fig \ref{schematic} $f)$, and used to check all line-plane crossings with the efficient \citet{Ramsey_bilinear_2004} algorithm.
The straight line to check is then between the current pair of flat surface crossings and patch $\xi^{BL}$ is shifted in multiples of $\Delta z$ to get the the successive bin surfaces between the $z$ coordinates of the surface crossings .
The pseudocode for this process is as follows,
\begin{lstlisting}
%\textcolor{function}{function}% surface_fluxes(r1, r2, quantity, fluxes)
  #Get all ordered crossing along line between r1 and r2
  crossings = get_flat_crossings(r1, r2)
  for i=1 to size(crossings)-1
    #Get cell indices for a pair of crossings
    cellA = get_intrinsic_cell(crossings(i))
    cellB = get_intrinsic_cell(crossings(i+1))
    #Check if same cell in both x and y 
    #which means separated by rows of cells 
    #as highlighted region in Fig %\textcolor{comment}{\ref{schematic} $f)$}%
    if (cellA(1) == cellB(1) & cellA(2) == cellB(2)
      #patch P is the z coordinate of the 
      #4 corners of all cells in row
      P = get_patch(cellA)
      #Loop in z direction between cellA and cellB
      for j=cellA(3) to cellB(3)
        #Get all crossing for line r1 + s*r12 and patch P
        #with j*dz shift to surface of cell j
        rc = line_patch_intersect(r1, r12, P+dz*j)
        #Use crossing location to assign to cell
        cellS = get_intrinsic_cell(rc)
        #Add flux quantity to array for mass, momentum or energy
        fluxes(cellS(1),cellS(2),cellS(3)) += quantity
      %\textcolor{keyword}{end for}% 
    %\textcolor{keyword}{end if}%
  %\textcolor{keyword}{end for}%
%\textcolor{function}{end function}% update_surface_fluxes
\end{lstlisting}

			% section{Implementation}
% !TEX root =  main.tex

% -----------------------------------------------------------------------------
\section{Results and Discussion}
 \label{sec:Results and Discussion}

In this section, the results for pressure on the moving surface are presented using the new surface flux equations derived in Sections \ref{sec:Theory}, summarised for implementation in \ref{sec:Implementation} with setup and methodology in \ref{sec:Methods}.
We start with a validation of the intrinsic density calculation and parameterise the bilinear approximation for varying resolutions.
Next, the novel surface flux forms are shown to be exactly conservative, before being compared to the volume averaged expressions.
The full balance of the surface flux contributions is shown next, highlighting the importance of surface evolution in the equations of motions and showing the normal pressure should be constant over the interface.
Finally, the calculation of the surface tension using the surface pressure is discussed.

\begin{figure}
  \centering
    \includegraphics[width=1.0\textwidth]{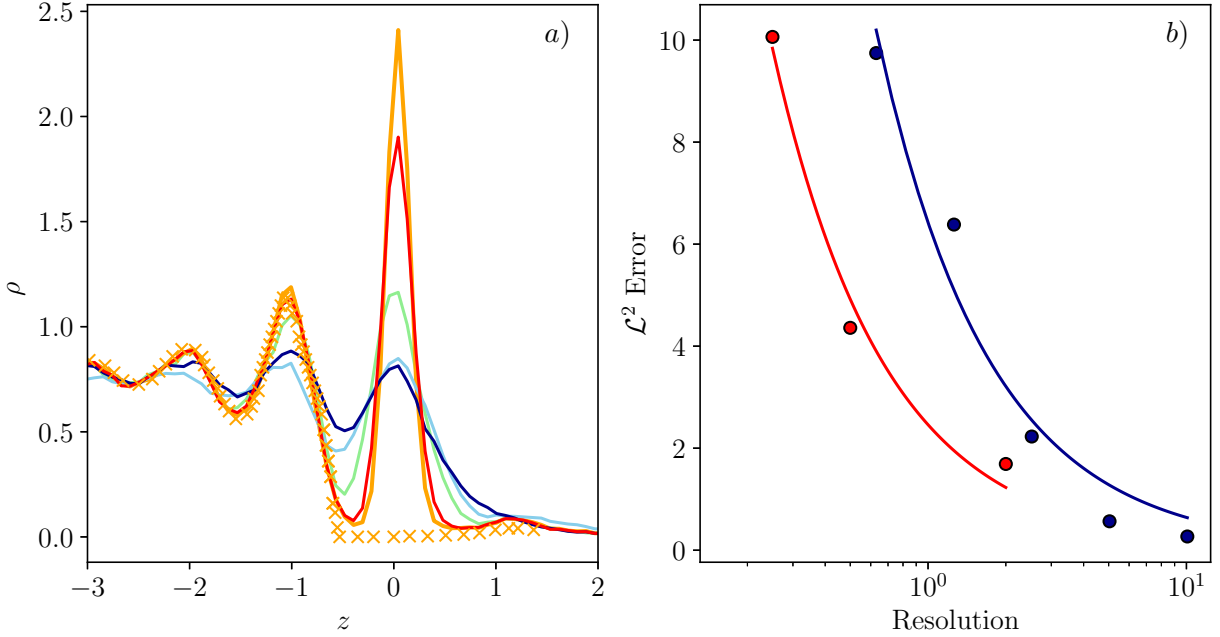}
\put(-260,230){$a)$}
\put(-25,230){$b)$}
      \caption{The effect on intrinsic density of the bilinear approximation is shown for different resolutions defined as $Resolution=\sigma/\Delta r$ which is the number of times the square bilinear patches of side length $\Delta r$ fit in an intermolecular spacing $\sigma$ and for intrinsic surfaces the number of wavelengths $\lambda_r$ per $\sigma$ with $Resolution=\sigma/ \lambda_r$. In $a)$ the full intrinsic solution is shown as a yellow line for the case where $\lambda_r = \lambda_{u}=\sigma$ compared to yellow crosses showing the solution from \citet{Chacon2003aa}. Bin $Resolution$ of $10$ and $5$ are omitted as indistinguishable from the yellow line, red is a $Resolution$ of $2.5$, green is $1.25$ and dark blue is $0.63$. The intrinsic interface with $\lambda_r=4$ is shown for comparison in light blue. In $b)$ the $L^2 \textrm{ Error}  = \int_{-6}^{2} \rho_{_{\Delta r = \sigma}} (z) - \rho_{_{\Delta r}}(z) dz$ obtained from the integral over $z$ from $-6$ which is has approximatly zero pressure to the other side of the interface at $2$ for both bilinear (blue) with binsizes $\Delta r=\{1.6, 0.8, 0.4, 0.2, 0.1\}$ and intrinsic (red) cases  $\lambda r=\{0.5, 2.0, 4.0\}$ are shown, where the $\lambda_{u}=1.0$ is defined to be zero error. The fitted lines are of the form $\mathcal{A}/Resolution$  with $\mathcal{A}=6.42$ for binsizes and $\mathcal{B}/Resolution$ with $\mathcal{B}=2.46$ for the intrinsic wavelength.}
      \label{Intrinsic_resolution}
\end{figure}

A study of bilinear resolutions is shown in Figure \ref{Intrinsic_resolution}, with the density profile used to assess the accuracy of the varying surface approximations.
This is calculated as described in section \ref{sec:Mapping} by mapping the molecular positions based on the intrinsic interface and then binning molecules to the appropriate volumes with width $\Delta z= 0.175$.
The density for the full intrinsic surface case is presented as a yellow line with $\lambda_r = \lambda_u = \sigma$ and compared to the results of \citet{Chacon2003aa} included as crosses in Fig \ref{Intrinsic_resolution} $a)$.
This surface with $\lambda_r = \sigma$ is then sampled to define a piecewise bilinear approximation of the surface.
The cases with $5$ and $10$ bilinear bins per $\sigma$ are indistinguishable from the intrinsic surface and are omitted in Fig \ref{Intrinsic_resolution} $a)$.
The case where each $\sigma$ unit has $2.5$ bins, $\Delta r = 0.4$, is shown as a red line, where despite some decrease in the peaks at zero, the density profile is largely identical. 
As increased sampling requires large memory requirements and slows calculation, the optimal case should provide good agreement for minimal resolution.
At $1.25$ bins per $\sigma$ shown in green on Figure \ref{Intrinsic_resolution} $a)$, the sampling clearly gives a smeared density peak and is deemed not sufficiently accurate to provide a good representation of the interface.
The case of $0.625$ bins is also shown in light blue with the intrinsic interface for $\lambda_r = 4 \sigma$ included for comparison in dark blue.
This highlights that poor binsize $Resolution$ has the same effect as a lower wavelength fitting of the intrinsic surface.
The $\mathcal{L}^2 \textrm{ Error}$ are shown in Figure \ref{Intrinsic_resolution} $b)$ defined as the absolute sum of the density binning obtained for a given resolution, minus the $\lambda_r = \sigma$ case at every bin between $-6$ and $2$.
The blue points are the varying bilinear cases, $\Delta r$, while the red points are different intrinsic minimum wavelength values, $\lambda_r$, including $\lambda_r = 2 \sigma$ and $\lambda_r = 4 \sigma$ giving a blurred density from underfitting as well as the case with $\lambda_r = 0.5 \sigma$ which, perhaps counter intuitively, gives a less sharp density profile due to overfitting.
The points for varying intrinsic and binsize resolution in Figure \ref{Intrinsic_resolution} $b)$ are reasonably well fitted by the lines which are of the form $\mathcal{A}/Resolution$ and $\mathcal{B}/Resolution$ with $\mathcal{A}=6.42$ and $\mathcal{B}=2.46$, suggesting error will tend to zero in the limit of infinite resolution but slowly.
Using bilinear approximations is therefore seen to main sufficient accuracy to obtain intrinsic quantities like density, with error over the whole plot negligibly small when $\Delta r < 0.2$ and reasonable with $\Delta r < 0.4$.
The effect of changing bilinear resolution can be seen to be similar to using a larger minimum wavelength when fitting the intrinsic surface, with roughly a two fold difference in trends ($\mathcal{A}/\mathcal{B} \approx 2.6$) so we need twice the increase in $\Delta r$ resolution to match a change in $\lambda_r$.
In this work, the presented plots use $\Delta r = 0.2 \sigma$ or $Resolution=5$ for the bilinear approximation to ensure measure quantities such as pressure are free from any artefacts.

\begin{figure}
  \centering
    \includegraphics[width=1.0\textwidth]{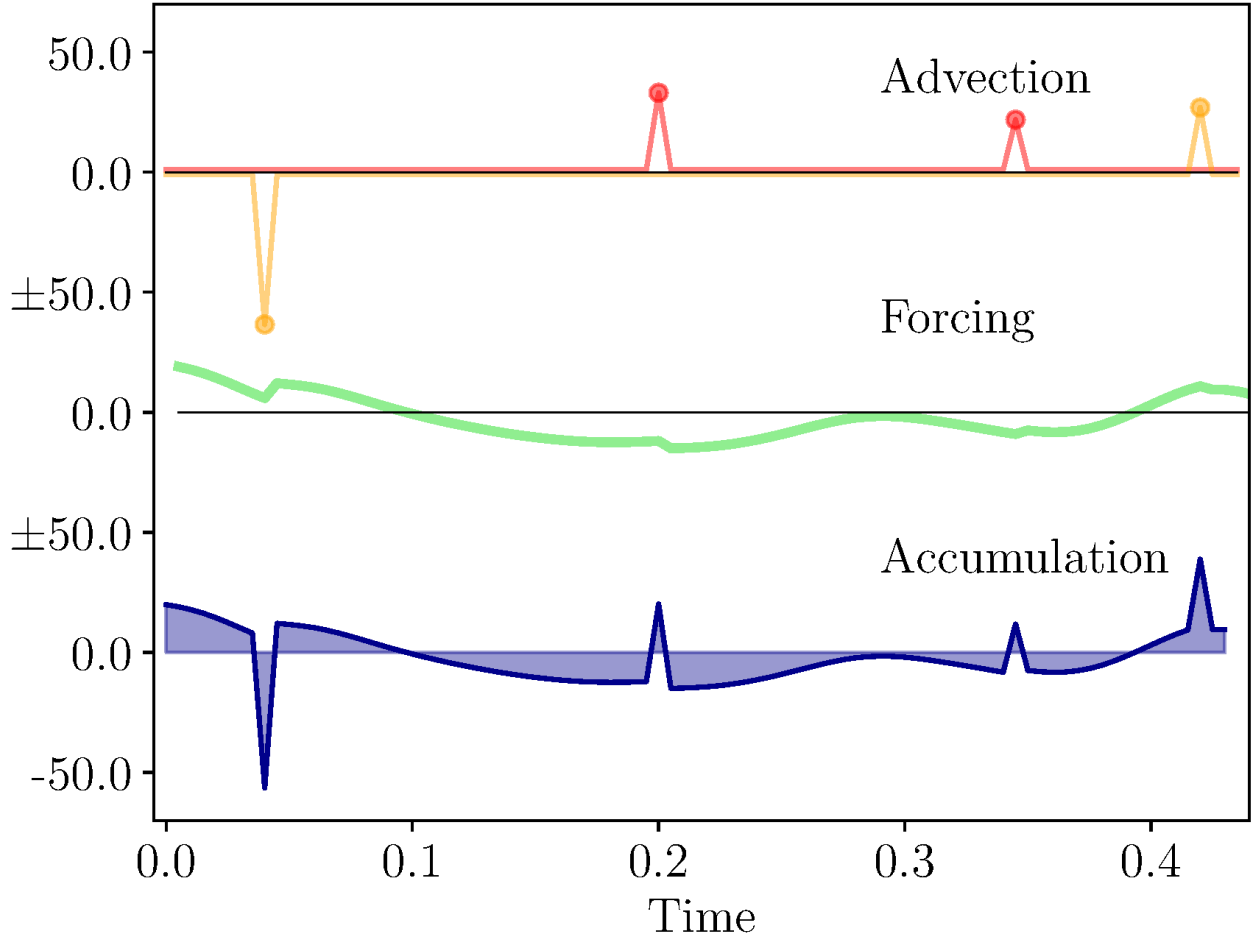}
      \caption{Components of momentum labelled in \eq{Labelled_CV_terms} for a control volume of width $\Delta z= 0.175$ with $x$ and $y$ extents of $3.97$, following the intrinsic surface with $Advection$ including fluxes of molecules over the surface (red) and molecule crossings due to surface movement (yellow), $Forcing$ the contributions due to forces between molecules crossing the volume surface (green) and $Accumulation$ the resulting change in momentum inside the control volume which is exactly equal to $Forcing + Advection$ ( checked to machine precision).}
      \label{CV_conservation}
\end{figure}

Having parameterised the effect of bilinear resolution, we move on to checking exact conservation for an arbitrary volume consisting of $10 \times 10$ bilinear segments of size $\Delta r =0.4$.
%The control volume form is unique in that it provides this exactly conservative form of the equations of motion.
The surface is therefore of size $\sim 4 \times 4$ in $x$ and $y$, centred on the intrinsic surface molecules, so it is the flux either side of the interface at $\pm \Delta z /2 = \pm 0.0875$ that is recorded, along with contributions on the top and bottom due to interactions with the remaining molecules on the intrinsic interface.
This is shown in Fig \ref{CV_conservation} where the terms in \eq{Labelled_CV_terms} are measured including the flux of molecules over the surface ($Advection$), both from molecules moving over the surface in red and the surface moving over molecules in yellow, plus the interaction force ($Forcing$) in green between the molecules all add up to the blue line for the change of momentum in the control volume ($Accumulation$).
The $Accumulation$ is shown in Fig \ref{CV_conservation} as a filled area under a curve to emphasise the fact that it is the integral of this area that determines the changing momentum inside the control volume.
This average momentum is constantly changing due to forces acting on the volume as well as molecules entering and leaving.
Exact conservation allows us to be sure that the implementation is correct, mollifies the issues associated with the non-uniqueness of the pressure tensor, providing an exact link between surface pressure on the volume and momentum change inside, as well as guaranteeing all possible terms which could contribute to surface tension have been included.
This exact conservation is valid for any arbitrary volume in space and is checked during all calculations in this work, providing a thorough validation of the mathematics and the implementation of the ray tracing on a complex moving surface.
The contributions due to forcing can be seen to appear as a continuous line in Fig \ref{CV_conservation}, as each molecule in the interface interacts with a neighbourhood of surrounding molecules within distance $r_c$.
The sum of all interactions on all molecules in the control volume, as well as between molecules on either side of the $\sim 4 \times 4$ patch, varies continuously as relative proximity changes.
The occasional large spikes due to molecules entering or leaving the volume represents an evaporation or condensation event.
These occur both due to molecular motion and when the fitting process of the interface finds a closer molecule than the current set, which can manifest in either the interface moving past the molecule or the molecule moving over the interface which is no longer following.
Molecules can also leave the volume in Fig \ref{CV_conservation} by diffusing along the intrinsic surface and leaving the $4$ by $4$ control volume.
This detailed balance has potential applications in designing improvements to the intrinsic fitting process, for example to minimise evaporation events or track surface transport.
The intrinsic fitting to a target density $\rho_s$ is equivalent to the mass control volume \eq{mass_CV_conservation} having zero $Accumulation$ and therefore ensuring zero net $Advection$ for a control volume the size of the whole interface.

\begin{figure}
  \centering
    \includegraphics[width=1.0\textwidth]{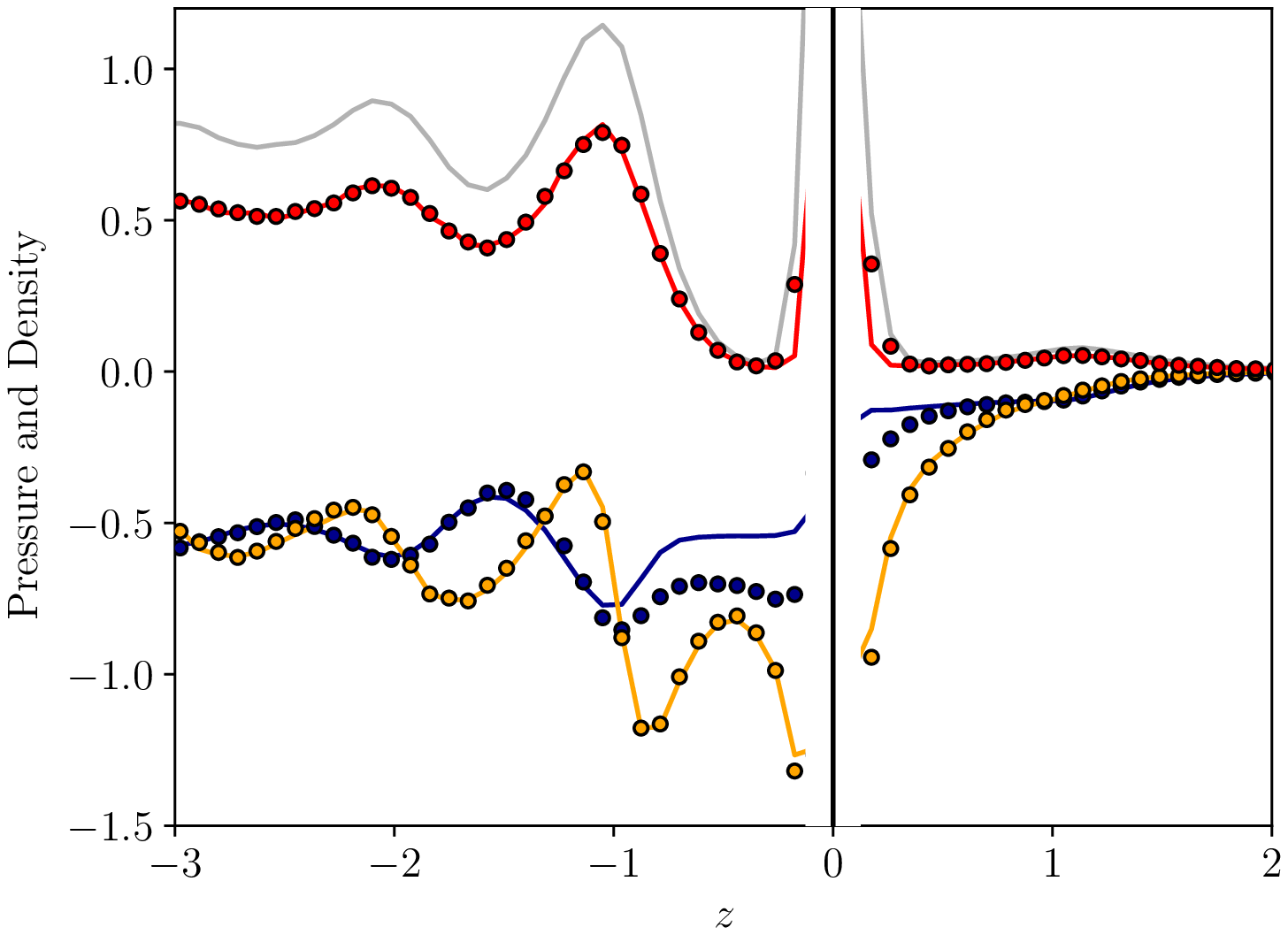}
      \caption{A comparison of pressure as a function of $z$ position, including surface tangential pressure \eq{y_equation_implementation} and surface normal pressure \eq{z_equation_implementation} (lines) compared to the volume average (points) pressure \eq{VA_relation} from previous work \citep{Braga_et_al18}. The kinetic pressure $\StressSurfscalar{\!}^{k^{^\prime}}_N$ and $\StressSurfscalar{\!}^k_T$ are shown as red lines, where the prime on the normal kinetic component denotes it does not include the surface movement, $\partial \xi / \partial t$, term (where kinetic normal $\StressSurfscalar{\!}^{k^{^\prime}}_N$ has an identical profile to the tangential component). Red points are $\PressureVAscalar{}^k_T$ and density of particles is shown as a grey line for reference. The tangential configurational pressure $\StressSurfscalar{\!}_T^c$ is shown as a yellow line with $\PressureVAscalar^c_T$ as yellow points and the normal component of configurational pressure $\StressSurfscalar{\!}_N^c$ is shown as a blue line with blue points for the VA term $\PressureVAscalar{}^c_N$. The position of the interface is plotted as a black line with a semi-transparent mask region to hide the peak at the intrinsic interface.}
      \label{ALL_plots}
\end{figure}

The new surface flux forms of pressure are compared to the volume average (VA) forms \eq{VA_relation} in Figure \ref{ALL_plots}. 
The volume average pressures are shown as points and surface pressures are shown by lines with kinetic pressure in red, tangential configurational pressure, $\StressSurfscalar{\!}^{c}_T \define \frac{1}{2} \left[\StressSurfscalar{\!}^{c}_{xx} + \StressSurfscalar{\!}^{c}_{yy}\right]$, shown in yellow and surface normal pressure  $\StressSurfscalar{\!}^{c}_N \define  \StressSurfscalar{\!}^{c}_{zz} $ in blue.
The kinetic pressure term of \eq{z_equation_implementation} is split into a surface evolution component $\partial \xi / \partial t$ and the remaining kinetic term denoted with a prime, $\StressSurfscalar{\!}^{k^{^\prime}}_N$, so that,
\begin{align}
\StressSurfscalar{\!}^{k}_N = \StressSurfscalar{\!}^{k^{^\prime}}_N  + \frac{\partial \xi }{ \partial t} %\textrm{  where   } \frac{\partial \xi }{ \partial t} = \frac{1}{\Delta S_z} \displaystyle\sum_{i=1}^N  m_i \dot{z}_i  \vartheta_t
\end{align}
and convection is assumed to be zero $\rho \boldsymbol{u} u_z = 0$.
The kinetic pressure calculated using the surface pressure definition is visually identical to the volume average one in Figure \ref{ALL_plots}. 
Note the normal component of surface kinetic pressure is shifted by $\Delta z/2$, as cell surfaces pressure is obtained on surfaces while volume average pressure is at the cell centres.
The density is shown in light grey for comparison, to highlight the kinetic pressure contributions are due to kinetic energy of molecules so directly correlated with their locations and identical in both surface and volume average measures.
The similarity in tangential components of configurational pressure, shown by the yellow points and lines, is consistent with past work \citep{Heyes_et_al}, which shows VA and MOP give identical results in the limit of small bins.
As the tangential contributions are calculated on a flat surface, using the form of \eq{y_equation_implementation} they are expected to be identical to the VA pressure expressions.
The normal components shown in blue is the only quantity that is seen to be different between surface and volume average pressure.
This is a direct result of the form of \eq{z_equation_implementation} which calculates the pressure dotted with the surface normal $\tilde{\boldsymbol{n}}_z$ at the location of every interaction for the interface at every time.
In contrast, the volume average tensor $f_{z ij} r_{z ij}$ is independent of the surface normal so remains aligned with the $z$ Cartesian axis.
This also makes clear why the tangential components are the same, the surface pressure is shown to be dotted with the $\boldsymbol{n}_y$ in the flat surface case which aligns with the volume average $f_{y ij} r_{y ij}$ components so giving identical results.
A similar observation is true for the other tangential component dotted with $\boldsymbol{n}_x$.

\begin{figure}
  \centering
    \includegraphics[width=1.0\textwidth]{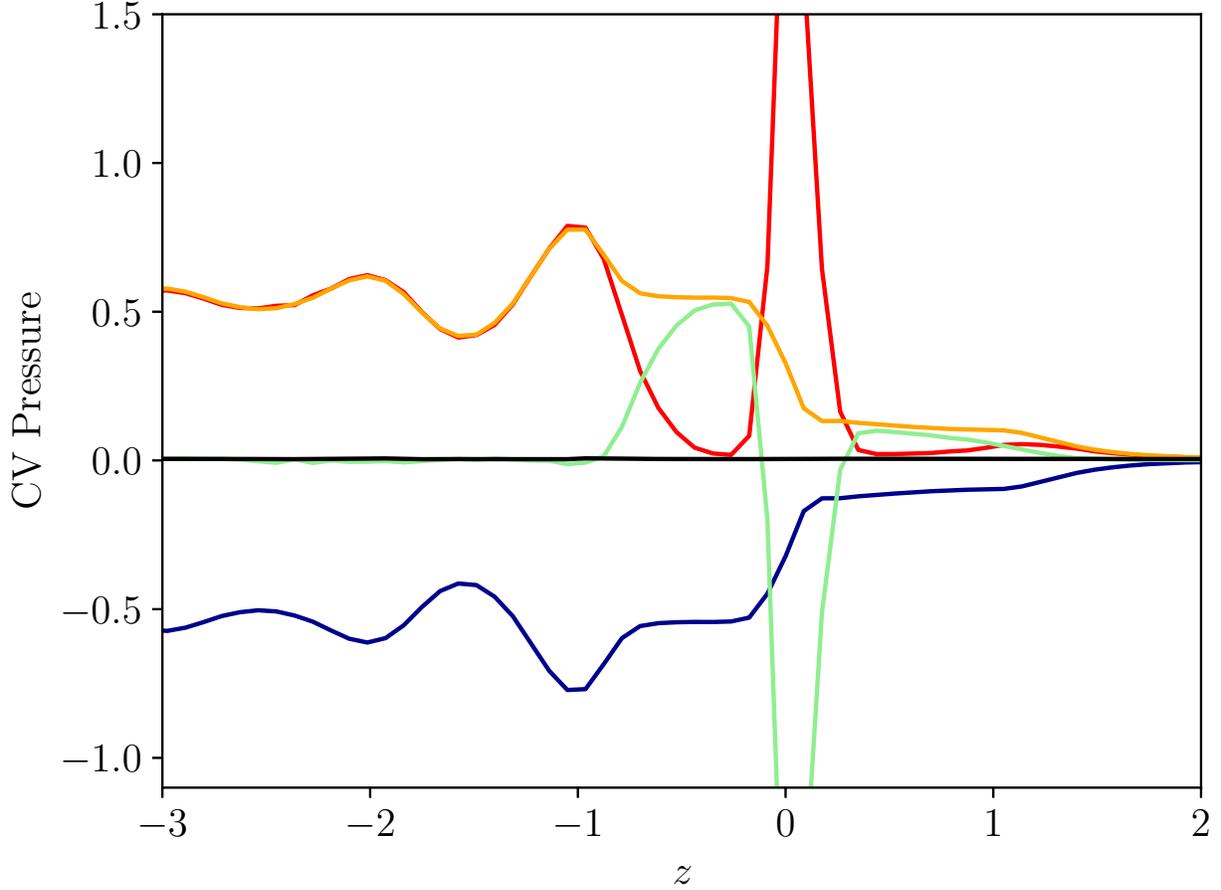}
      \caption{All terms which contribute to the normal component of the total CV Pressure including configurational pressure $\StressSurfscalar{\!}_N^c$ shown in blue, kinetic pressure  $\StressSurfscalar{\!}_N^{k^{^\prime}}$ shown in red and surface movement $\partial \xi / \partial t$ shown in green, with the total kinetic contribution $\StressSurfscalar{\!}_N^k =  \StressSurfscalar{\!}_N^{k^{^\prime}} + \partial \xi / \partial t$ shown in yellow. The total pressure $\StressSurfscalar{\!}_N = \StressSurfscalar{\!}_N^k+\StressSurfscalar{\!}_N^c$, black line, is a small constant value over the surface as required to ensure the surface is not moving. }
      \label{CV_terms}
\end{figure}

Using surface pressure normal to the instantaneous surface can be shown to be essential to the exact balance of momentum over the surface.
To see this, in Figure \ref{CV_terms} all contributions to the surface in \eq{z_equation_implementation} are plotted on the same graph.
The configurational (blue, $\StressSurfscalar{\!}_N^c$) and kinetic (red, $\StressSurfscalar{\!}_N^{k^{^\prime}}$) pressure terms are identical to the ones plotted on figure \ref{ALL_plots}, but when the surface fluctuations (green, $\partial \xi / \partial t$) are added to give total kinetic pressure (yellow, $\StressSurfscalar{\!}_N^{k}$), it can be seen that the resulting profile perfectly mirrors the configurational pressure.
The result is the sum of the extended set of kinetic terms and configurational pressure gives a perfectly flat normal pressure over the surface, a required result for the interface to be stationary.
Demonstration of a constant pressure profile near a wall in a molecular system was shown to be an important reason for using the VA or MOP form of pressure instead of the virial or IK1 expressions \citep{Todd_et_al_95, Heyes_et_al}.

The shape of the configurational pressure profile $\StressSurfscalar{\!\!}_N^c$ has a flat region from $z=0.2$ to $0.8$ before the first liquid peak at $z \approx 1$.
The total kinetic part, $\StressSurfscalar{\!\!}_N^{k}$, including $\partial \xi / \partial t$, in yellow exactly mirrors the configurational with this same flat region.
It can be seen this flat region, that is, $\StressSurfscalar{\!}_N^{k^{^\prime}} + \partial \xi / \partial t = Constant$ from $z=0.2$ to $0.8$ results from an exchange between contributions due to surface movement and contributions due to flux of molecules as we move in the $z$ direction.
This is a consequence of the intrinsic surface fitting process, by choosing our reference frame to follow the interface molecules the measured kinetic pressure exposes the liquid structure in a similar way to the radial distribution function, in particular the liquid's tendancy to separate between molecular layers results in a drop to zero in the gap between the interface and first fluid layer.
The interface movement term $\partial \xi / \partial t$ captures the movement of the intrinsic interface, a reference frame which tracks the surface molecules, and that allows the plot to identify the liquid structure peaks observed 
Put another way, the $\Pi^{k^{^\prime}}$ term captures the structure inside the moving liquid cluster interface, the $\partial \xi / \partial t$ captures how that structure moves. 

It is worth noting that Figure \ref{CV_terms} shows the common equilibrium assumption $\boldsymbol{\nabla} \cdot \Pi = 0$ applied in statistical mechanical derivation of surface tension \citep{Rowlinson93, Malijevsk__2012} is only valid if the contribution from surface movement and local curvature are correctly included.
The VA terms in Fig \ref{ALL_plots} will not satisfy this condition due to missing curvature term shown in Appendix \ref{sec:surface_terms}.
This may suggest the equilibrium assumption is a source of error for spherical and cylindrical volumes. A full mechanical approach may provide insight by including kinetic fluxes, surface movement and changing momentum of the volume as shown in Figure \ref{CV_conservation}.

\begin{figure}
  \centering
    \includegraphics[width=1.0\textwidth]{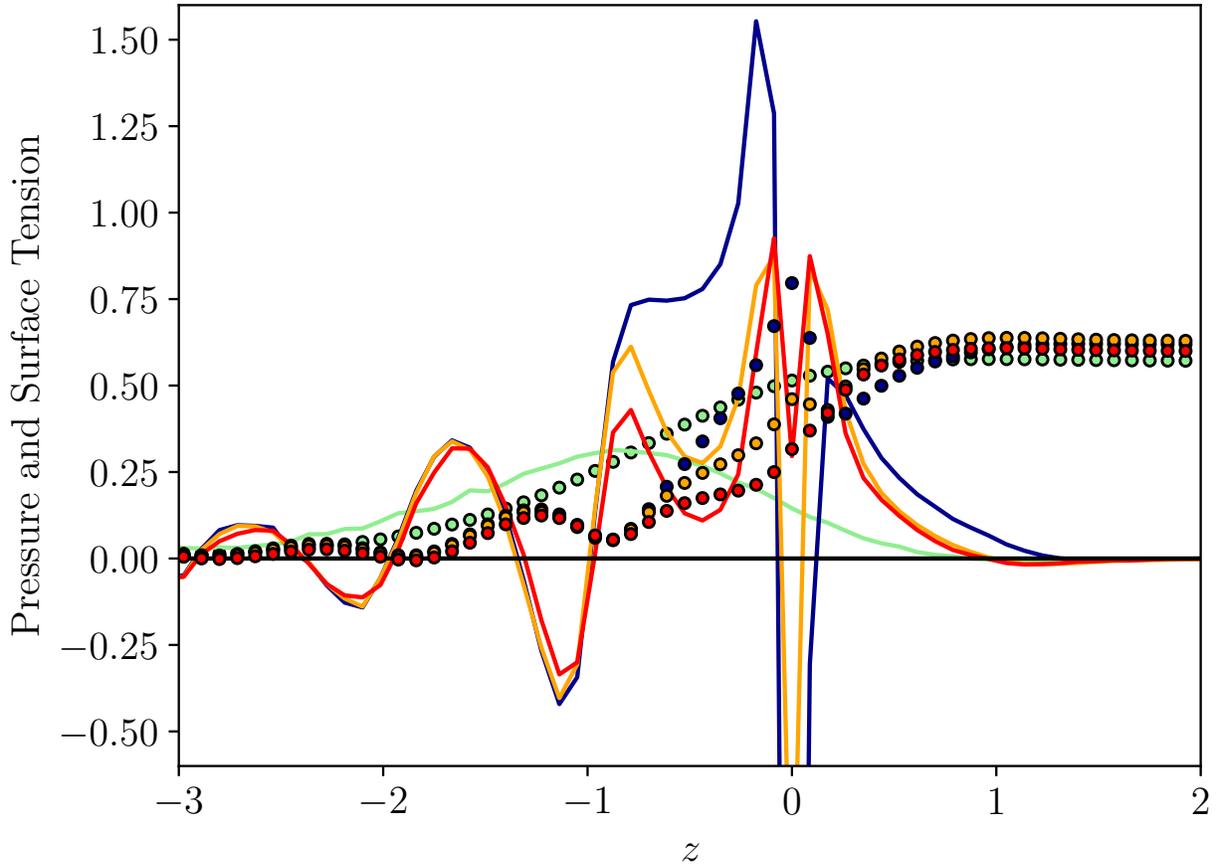}
      \caption{Calculation of the surface tension with normal minus tangential pressure $\Pi_N - \Pi_T$ as red lines for the full VA pressure, yellow lines for the surface configurational part only and blue lines the entire surface pressure including the surface evolution, where the minimum at $z=0$ of the yellow and blue lines are not shown as they go to $-1.2$ and $-4.9$ respectively. The corresponding surface tension calculated using the MOP on a fixed grid is shown in green for reference, shifted so zero is the location where $\frac{1}{2}(\rho_l + \rho_g)$. The corresponding cumulative integral of each curve to a given $z$ value, $\gamma (z) = \int_{-6}^{z} \left[\Pi_N(z^{\prime}) - \Pi_T(z^{\prime}) \right] dz^{\prime}$, is shown in the same colour, with red circles VA pressure, yellow circles configurational surface, blue circles the full surface pressure and green the fixed grid MOP pressure.}
      \label{surface_tension}
\end{figure}

Finally, we consider the surface tension in Figure \ref{surface_tension}.
In order to explore the various contributions to surface tension from different parts of pressure, three different Kirkwood Buff  \citep{Kirkwood_Buff} style formula are plotted,
 \begin{align}
\SurfaceTensiontype\limits^{\scriptscriptstyle{V\!A}} & = \int_{-6}^{z} \left[\PressureVAscalar\! {_{N}} - \PressureVAscalar\! {_{T}} \right] dz^\prime \label{VA_ST}
  \\
\SurfaceTensiontype\limits^{\scriptscriptstyle{Surf}}{}^c(z) & = \int_{-6}^{z} \left[\StressSurfscalar\! {^c_{N}} - \StressSurfscalar\! {^c_{T}} \right] dz^\prime \label{CVc_ST}
  \\
\SurfaceTensiontype\limits^{\scriptscriptstyle{Surf}}(z) & = \int_{-6}^{z} \left[\StressSurfscalar\! {_{N}} - \StressSurfscalar\! {_{T}} \right] dz^\prime 
\approx \SurfaceTensiontype\limits^{\scriptscriptstyle{Surf}}{}^c(z) + \int_{-6}^{z} \frac{\partial \xi(z^\prime) }{ \partial t } dz^\prime 
%\approx \left< \StressSurfscalar\! {_{N}} \right> (z+3)- \int_{-6}^{z} \StressSurfscalar\! {_{T}} dz^\prime
 \label{CV_ST}
 \end{align}
which includes the surface tension from the volume average pressure \eq{VA_ST}, the configuration part of the surface pressure \eq{CVc_ST} and from the full surface pressure \eq{CV_ST}.
In previous work \citep{Braga_et_al18}, the contributions to surface tension from the volume average terms were discussed, and the red lines and circles in Figure \ref{surface_tension} are presented as a basis for comparison, together with the method of planes pressure measurements obtain using a fixed grid shown by green lines and circles.
The configurational part is important as the normal component of configurational surface pressure was the only term which showed a difference when compared to the volume average pressure in Figure \ref{ALL_plots}.
As the kinetic normal and tangential components are identical for both volume average and surface pressure these cancel in the \citet{Kirkwood_Buff} formula and the configurational terms contain all contributions to surface tension.
As a result, any difference between the surface tension calculated from the VA pressure and surface pressure would be expected to come from the $\StressSurf\! {^C_{N}}$ term. 
The red and yellow lines in Figure \ref{surface_tension} show the difference between these two pressure measurements is mostly located between the intrinsic interface and first layer in the liquid region from $z=0$ to $1$.
The resulting integral to give surface tension from \eq{CVc_ST} (yellow circles) can be seen to still converge to the same value as the volume average \eq{VA_ST} (red circles) in Figure \ref{surface_tension} (and the surface tension obtained from the fixed references frame pressure shown by green circles).
However, the surface pressure sees a greater contribution to surface tension in the liquid region, from $z=-0.5$ to $z=0$, than the VA. 
This is offset by a much larger negative contribution from the intrinsic surface molecules themselves at $z=0$.
This is more pronounced for the full surface pressure, the blue line, and resulting surface tension using \eq{CV_ST}, shown by blue circles in Figure \ref{surface_tension}.
There is an almost linear contribution from $z=-1$ to $z=0$ and an even larger negative contribution from the interface molecules at $z=0$ when compared to the configurational part of \eq{CVc_ST}.
The interface control volume sits on the interfacial molecules themselves, so this peak represents the tangential forces between them, as well as a contribution due to the movement of the surface. 
The key difference between the configurational \eq{CVc_ST} and the full surface pressure \eq{CV_ST} is the inclusion of surface evolution in time $\partial \xi / \partial t $.
This can be seen to give a net zero contribution to surface tension in this equilibrium case, but redistributes giving a large negative contribution from the surface itself at $z=0$ and equal positive contribution between $z = \pm 1$ in both the liquid and vapour region.

The interface tracking surface pressure derived in this work gives further insight into the surface tension distribution, while still integrating to the same overall value as obtained by a fixed reference frame.
However, the simulation of a flat equilibrium interface presented here is the simplest possible case.
The real strength of the derived formulation is that it is valid away from equilibrium for any interface described by a function of the form $\xi(x,y,t)$, with the control volume formulation providing exact conservation every single timestep.
This exact momentum balance on a complex time evolving interface could provide useful insights in bubble nucleation, film rupture or contact line dewetting. 
%The consequence of the normal pressure being constant in $z$ is that it only contributes a constant $\mathcal{A}$ to the surface tension calculation of \eq{CV_ST}.
% 
% \textbf{NOTE - why are VA and surface N-T different but integral the same. Normal irrelevant in VA? If kinetic is essential for flat normal, it then contributes to ST in tangential case as no longer simply cancels}
% As the kinetic part is identical in the normal and tangential direction, it does not contribute in the mechanical definition of surface tension from the formula of \citet{Kirkwood_Buff}.
% However, as it cancels the configurational part in the surface form, it once again becomes important whereas the normal component becomes a constant contribution, 
%  \begin{align}
%  \gamma = \int_{-\infty}^{\infty} \left[\Pi_N - \Pi_T \right] dz \approx \mathcal{C}- \int_{-\infty}^{\infty} \Pi_T dz
%  \end{align}
% where $\mathcal{C}$ is a constant from the average normal pressure.

			% section{Results}

\section{Conclusions}
\label{sec:Conclusions}

In this work, we derive a new formulation of surface pressure in a reference frame moving with the interface between a liquid and a vapour.
This derivation starts from statistical mechanical definitions \citep{Irving_Kirkwood} of density, momentum and energy equations but without ensemble average.
The resulting equations are integrated over a volume fitted to the interface as it evolves in time, giving instantaneous surface fluxes on the moving surface.
These surface flux equations are shown to include two extra contributions, one due to the instantaneous surface curvature at the point of surface crossings and one due to the movement of the surface in time.
By including all contributions for curvature and surface movement, the equations can be shown to be exactly conservative in a molecular dynamics (MD) simulation.
These extra terms are also shown to be essential to provide an exact force balance over the moving liquid-vapour interface. 
The derived equations are presented in a form which can be easily implemented in MD, applying a ray tracing approach commonly used for computer graphics.
Several insights into the pressure and surface tension are presented for the simplest case of a flat interface between the liquid and vapour.
Despite being tested in a simple system, the derived equations make no assumption other than mass conservation and Newton's laws.
As a result, they provide exact expressions for the equations of hydrodynamics which can take any local region on the surface and follow the instantaneous surface shape as it evolves.
Therefore, the new equations are expected to have great potential applications to understand a range of hydrodynamics phenomena including Marangoni effects, growing bubbles, moving contact lines and deforming interfaces.

%Mechanical descriptions of an interfacial region usually rely on an expression for the pressure tensor across a planar surface, as described by Kirkwood and Buff [J. Chem. Phys., 17(3), (1949)], and later Irving and Kirkwood [J. Chem. Phys. 18, 817 (1950)].
%Our previous work [J. Chem. Phys. 149, 044705 (2018)] relaxed this condition and generalised Irving and Kirkwood’s expression for the pressure tensor [J. Chem. Phys. 18, 817 (1950)] to fluctuating, non planar surfaces free from the thermal smearing of capillary waves.

%
%
%Local form to get changes at any point on the moving surface. \\
%
%This provides a detailed insight into the interface, with every contribution to the dynamics guaranteed by the momentum balance.
%
%Need exact form to be valid instantaneously and arbitrarily far from equilibrium. \\

\section{Data Availability Statement}

The data that support the findings of this study are available from the corresponding author upon reasonable request.

% !TEX root =  main.tex

% -----------------------------------------------------------------------------

\appendix
\section{Integrating the Equations}
\label{sec:Appendix_CV}

This appendix details the full mathematics of the process used to provide the equations for use in a molecular dynamics simulation.
First to highlight the similarity between the configurational and kinetic part we integrate the kinetic expression over time in \eq{z_stress}.
This can be interpreted in two ways, either as part of time averaging or as part of the evolution of the system,
  \begin{align}
\int_{t_1}^{t_2} \left[ \rho \boldsymbol{u} u_z + \boldsymbol{\Pi}_z^k \right] dt
 = \frac{1}{\Delta S_z}  \displaystyle\sum_{i=1}^N  m_i \boldsymbol{\dot{r}}_i & \int_{t_1}^{t_2}  \Big [\dot{x}_i \frac{\partial \insurfi{+}}{\partial x_i} + \dot{y}_i \frac{\partial \insurfi{+}}{\partial y_i}  + \dot{z}_{i}
+  \frac{\partial \insurfi{+}}{\partial t} \vphantom{\frac{\partial \insurfi{+}}{\partial y_i}} \Big] dS_{zi}^+ dt
\nonumber  \\ 
\boldsymbol{\Pi}_z^c = \frac{1}{2}  \frac{1}{\Delta S_z}\sum_{i,j}^{N} \boldsymbol{f}_{ij}  
& \int_0^1 \Big[  {x}_{ij} \frac{\partial \insurfl{+} }{\partial x_\lambda} + {y}_{ij} \frac{\partial \insurfl{+} }{\partial y_\lambda} + {z}_{ij}  \Big]  dS_{z\lambda}^+  d\lambda .
\label{Full_compared_stresses}
 \end{align}
The expression for pressure in \eq{Full_compared_stresses} can then be written as follows,
  \begin{align}
\int_{t_1}^{t_2}  \left[ \rho \boldsymbol{u} u_z + \boldsymbol{\Pi}_z^k \right] dt
 = \frac{1}{\Delta S_z}  \displaystyle\sum_{i=1}^N  m_i \boldsymbol{\dot{r}}_i \Bigg( & \int_{0}^{1}  
 \Big [\frac{dx_\tau}{d\tau} \frac{\partial \insurfi{+}}{\partial x_i} 
     + \frac{dy_\tau}{d\tau} \frac{\partial \insurfi{+}}{\partial y_i}  
     + \frac{dz_\tau}{d\tau}
\vphantom{\frac{\partial \insurfi{+}}{\partial y_i}} \Big] dS_{zi}^+ d\tau + \vartheta_t \Bigg)
\nonumber  \\ 
\boldsymbol{\Pi}_z^c = \frac{1}{2} \frac{1}{\Delta S_z}\sum_{i,j}^{N} \boldsymbol{f}_{ij}  
& \int_0^1 
\Big[  \frac{\partial x_{\lambda}}{\partial \lambda} \frac{\partial \insurfl{+} }{\partial x_\lambda} 
     + \frac{\partial y_{\lambda}}{\partial \lambda} \frac{\partial \insurfl{+} }{\partial y_\lambda} 
     + \frac{\partial z_{\lambda}}{\partial \lambda}  \Big]  dS_{z\lambda}^+  d\lambda,
\label{integrated_kinetic_vs_config}
 \end{align}
where we use the substitution $\tau = [t - t_1]/[t_2 - t_1]$ in the integral over time to get the same form as the interaction over path $\lambda$, denoting $\boldsymbol{r}_1 \define \boldsymbol{r}_i(t_1) $ and $ \boldsymbol{r}_2 \define \boldsymbol{r}_i(t_2) $ so $\boldsymbol{r}_i(t) = \boldsymbol{r}_\tau = \boldsymbol{r}_1 + \tau \boldsymbol{r}_{12}$ with $\boldsymbol{r}_{12} = \boldsymbol{r}_2 - \boldsymbol{r}_1$ and $\tau$ from $0$ to $1$.
Now both expressions are in the form $\boldsymbol{r}_\alpha =  \boldsymbol{r}_1 + s \boldsymbol{r}_{12}$ with $\alpha=\{ \tau, \lambda \}$ and $s=\{\tau, \lambda$\} allowing us to write the generalised control volume function, 
\begin{align}
\vartheta_s =  
 &  \left[ H \left( x^+ - x_\alpha(s)  \right)  -  H \left( x^- - x_\alpha(s)  \right)\right] \;\;\;\;
  \nonumber \\ 
 \times &  \left[ H \left( y^+ - y_\alpha(s)  \right)  -  H \left( y^- - y_\alpha(s)  \right)\right] 
\nonumber \\ 
\times &  \left[ H \left( \insurfi{+}  - z_\alpha(s)   \right) -  H \left( \insurfi{$-$} - z_\alpha(s)   \right) \right] 
= \Lambda_x (s) \Lambda_y (s) \tilde{\Lambda}_z (s),  \;\;\;\;
\label{general_CV}
 \end{align}
where each directional difference between two Heaviside functions is written using shorthand $\Lambda_\beta$ with $\beta=\{x,y,z\}$ and just the functional dependence on $s$ shown.
The tilde on the $\Lambda_z$ term denotes the intrinsic interface component is included on this surface.
Both integral expressions in \eq{integrated_kinetic_vs_config} are in the form of an integral of the derivative of \eq{general_CV} with respect to $s$.
More generally, the total expression for all surfaces can be written concisely as,
\begin{align}
 \int_{0}^{1} \frac{\partial \vartheta_s}{\partial s} ds 
=  \int_{0}^{1} \frac{\partial}{\partial s} \left[ \Lambda_x (s) \Lambda_y (s) \tilde{\Lambda}_z (s) \right] ds 
=   \int_{0}^{1} \left[ 
	 \frac{\partial \Lambda_x}{\partial s} \Lambda_y \tilde{\Lambda}_z 
+    \Lambda_x  \frac{\partial \Lambda_y}{\partial s}  \tilde{\Lambda}_z
+  \Lambda_x \Lambda_y \frac{\partial  \tilde{\Lambda}_z }{\partial s} \right] ds ,
\label{dCV_general}
 \end{align}
for example, with $s=\lambda$ the last term describes the difference between top and bottom configurational term in $z$,
\begin{align}
 \int_{0}^{1} \Lambda_x \Lambda_y \frac{\partial  \tilde{\Lambda}_z }{\partial s}  ds =  \int_{0}^{1}
 \Big[  \frac{\partial x_{\lambda}}{\partial \lambda} \frac{\partial \insurfl{+} }{\partial x_\lambda} 
     + \frac{\partial y_{\lambda}}{\partial \lambda} \frac{\partial \insurfl{+} }{\partial y_\lambda} 
     + \frac{\partial z_{\lambda}}{\partial \lambda}  \Big]  dS_{z\lambda}^+  
- \Big[  \frac{\partial x_{\lambda}}{\partial \lambda} \frac{\partial \insurfl{-} }{\partial x_\lambda} 
     + \frac{\partial y_{\lambda}}{\partial \lambda} \frac{\partial \insurfl{-} }{\partial y_\lambda} 
     + \frac{\partial z_{\lambda}}{\partial \lambda}  \Big]  dS_{z\lambda}^-  d\lambda.
 \end{align}

%where the intergal limits are just denoted as $s^-$ and $s^+$ which can be $0$ to $1$ for intermolecular interaction path or $t_1$ to $t_2$ for time evolution.
For the $x$ and $y$ surfaces, the intersection of a plane and line can be obtained directly, considering the first term on the right-hand side of \eq{dCV_general},
\begin{align}
 \int_{0}^{1}  \frac{\partial \Lambda_x}{\partial s} \Lambda_y \tilde{\Lambda}_z ds =  \int_{0}^{1} \frac{d x_\alpha } {ds} \left[ \delta \left( x^+ - x_\alpha(s)  \right)  -  \delta \left( x^- - x_\alpha(s)  \right)\right] \Lambda_y(s) \tilde{\Lambda}_z(s) ds.
 \end{align}
We apply the property of the Delta function to express it as the sum of its roots,
\begin{align}
 \delta \left( x^+ - x_\alpha   \right) = \displaystyle\sum_{k=1}^{N_{roots}} \frac{\delta \left( s - s_{k} \right)}{| d x_\alpha(s_k) / ds  |},
\label{DD_roots}
\end{align}
so positive surface in $x$ can then be expressed,
\begin{align}
 \int_{0}^{1} \frac{\partial x_\alpha (s) } {\partial s}  \delta \left( x^+ - x_\alpha(s)  \right)  \Lambda_y(s) \tilde{\Lambda}_z(s) ds 
=  \int_{0}^{1} \frac{\partial x_\alpha(s) } {\partial s}  \displaystyle\sum_{k=1}^{N_{roots}} \frac{\delta \left( s - s_{k} \right)}{| d x_\alpha(s_k) / ds  |}   \Lambda_y(s) \tilde{\Lambda}_z(s) ds 
\nonumber \\
=  \underbrace{\frac{ \partial x_\alpha(s_k) / \partial s  }{|\partial x_\alpha(s_k) / \partial s |}}_{i) \textrm{ Direction of crossing}}
 \displaystyle\sum_{k=1}^{N_{roots}} \underbrace{\left[ H \left( 1 - s_{k} \right) - H \left( - s_{k} \right) \right] \vphantom{\left( \frac{\partial x_\alpha } {\partial s}  \right)}}_{ii)\textrm{ Crossing between limits}}
 \underbrace{\Lambda_y(s_k) \tilde{\Lambda}_z(s_k) \vphantom{\left( \frac{\partial z_\alpha } {\partial s}  \right)} }_{iii)\textrm{ Crossing on yz CV Surface}}.
\label{flat_surface}
\end{align}
The three annotated terms include $i)$ a signum function which determines the direction of crossing with 
\begin{align}
 \frac{ \partial x_\alpha(s_k) / \partial s  }{|\partial x_\alpha(s_k) / \partial s |} = \frac{x_{12}}{|x_{12}|} = \frac{\boldsymbol{n}_x \cdot \boldsymbol{r}_{12} }{ |\boldsymbol{n}_x \cdot \boldsymbol{r}_{12}|},
\end{align}
using $d x_\alpha / ds = x_{12}$ so $i)$ is expressed in terms of the normal to the $x$ surface $\boldsymbol{n}_x = [1,0,0]$.
For part ii) of \eq{flat_surface}, the Heaviside functions between integral limits is only non-zero if the point is on the line between $r_1$ and $r_2$.
The expression for this root $s_k$ on the flat surfaces is the intersection of a plane and a line \citep{Smith_et_al12}, obtained by equating the surface $x^+$ and line $x_1 + s x_{12}$ to solve for the value of $s$ at crossing $ s_k =  (x^+- x_1)/x_{12} \define x_k^+$ allowing the crossing between limits term of \eq{flat_surface} to be expressed as,
\begin{align}
\left[ H \left( 1 - s_{k} \right) - H \left(  - s_{k} \right) \right]   
 = H \left( \frac{x^+- x_2}{x_{12}} \right) - H \left( \frac{x_1- x^+}{x_{12}} \right) 
 \nonumber \\
 = \frac{1}{2} sgn\left(\frac{1}{x_{12}}\right) \left[sgn(x^+- x_2) - sgn(x_1- x^+)\right],
\label{MOP}
\end{align}
which is the expression found in the method of planes form of stress and determines if $r_1$ and $r_2$ are on opposite sides of the plane \citep{Todd_et_al_95}.
% Finally, iii) the crossing on the flat $x^+$ surface is checked by the $\Lambda$ functions to ensure it is between the limits of the control volume surface in $y$ and $z$, where the $z$ varies as the intrinsic surfaces moves.
Finally, iii) the location of the crossing $s_k$ on the flat $x^+$ surface gives a value of zero for $\Lambda(s_k)$ when not within the limits of the control volume surface in the $y$ and $z$ directions, where the $z$ CV surface moves as the intrinsic surfaces moves, shown in  Fig \ref{CV_function}.
The location of crossing in each direction is,
\begin{align}
 x_\alpha(s_k) = x^+; \;\;\; y_\alpha(s_k) =  y_1 + s_k y_{12}; \;\;\; z_\alpha(s_k) =  z_1 + s_k z_{12},
\end{align}
and the $\tilde{\Lambda}_z(s_k)$ function is,
\begin{align}
\tilde{\Lambda}_z(s_k) %\;\;\; &  H \left( z^+ + \xi \big(x_\alpha(s_k),y_\alpha(s_k) \big) - z_\alpha(s_k)   \right)
%-  H \left( z^- + \xi \big(x_\alpha(s_k),y_\alpha(s_k) \big)  - z_\alpha(s_k)   \right) 
%\nonumber \\
= \;\;\; &  H \left( z^+ + \xi \left(x^+, y_1 + y_{12}s_k \right)  - z_1 -  \ z_{12} s_k  \right)
\nonumber  \\  
- & H \left( z^- + \xi \left(x^+, y_1 + y_{12}s_k \right)  - z_1 -  \ z_{12} s_k   \right) .
\label{MOP_local}
\end{align}
% with crossing locations on the $x^+$ plane of $x_\alpha(s_k) = x^+$ and $y_\alpha(s_k) =  y_1 + x_k^+ y_{12}$ used in the $\xi$ function to obtain the intrinsic surface value at the location of the crossing. The entire $\tilde{\Lambda}_z$ function then tests if the $z$ crossing location $z_\alpha(s_k) =  z_1 + x_k^+ z_{12}$ is between the limits $z^- + \xi$ and $z^+ + \xi$ can will be counted as a flux on that CV surface, as shown in Fig \ref{CV_function}.
The stress on the $x^+$ surface is therefore,
  \begin{align}
\int_{t_1}^{t_2}  \left[ \rho \boldsymbol{u} u_x + \boldsymbol{\Pi}_x^k \right] dt = & \frac{1}{\Delta S_x}  \displaystyle\sum_{i=1}^N  m_i \boldsymbol{\dot{r}}_i  
 \frac{ \boldsymbol{r}_{i_{12}} \cdot \boldsymbol{n}_x }{|\boldsymbol{r}_{i_{12}} \cdot \boldsymbol{n}_x |} 
\! \left[H \left( \frac{x^+- x_{i_2}}{x_{i_{12}}} \right) - H \left( \frac{x_{i_1}- x^+}{x_{i_{12}}} \right) \right] \Lambda_y(t_k) \tilde{\Lambda}_z(t_k) 
  \nonumber \\
\boldsymbol{\Pi}_x^c = & \frac{1}{2} \frac{1}{ \Delta S_z}\sum_{i,j}^{N} \boldsymbol{f}_{ij}  
\frac{\boldsymbol{{r}}_{ij} \cdot {\boldsymbol{n}}_x }{|\boldsymbol{r}_{ij} \cdot {\boldsymbol{n}}_x|} 
\; \left[H \left( \frac{x^+- x_j}{x_{ij}} \right) - H \left( \frac{x_i- x^+}{x_{ij}} \right) \right] \Lambda_y(\lambda_k) \tilde{\Lambda}_z(\lambda_k),
\nonumber
 \end{align}
note the inclusion of the index $i$ so $\boldsymbol{r}_{12} \to \boldsymbol{r}_{i_{12}}$ to emphasise it is molecule $i$ which is evolving from time $t_1$ to $t_2$ 
The link to the common method of planes (MOP) expression in the literature can be seen using $x/|x| = sgn(x)$ and \eq{MOP} to give, 
  \begin{align}
 \frac{ \boldsymbol{r}_{12} \cdot \boldsymbol{n}_x }{|\boldsymbol{r}_{12} \cdot \boldsymbol{n}_x |} 
\! \left[H \left( \frac{x^+- x_2}{x_{12}} \right) - H \left( \frac{x_1- x^+}{x_{12}} \right) \right]  \;\;\;\;\;\;\;\;\;\;\;\;\;\;\;\;\;\;\;\;\;\;\;\;\;\;\;\;\;\;\;\;\;\;\;\;\;\;\;\;\;\;\;\;\;\;\;\;\;\;\;\;\;\;\;\;\;\;\;\;\;\;\;\;
\nonumber \\
=  \frac{1}{2} sgn(x_{12}) sgn\left(\frac{1}{x_{12}}\right) \left[sgn(x^+- x_2) - sgn(x_1- x^+)\right]
  \nonumber \\
=  \frac{1}{2} \left[sgn(x^+- x_2) - sgn(x_1- x^+)\right]
 \end{align}
so the expressions for stress on the $x$ surface are the well know MOP form,
  \begin{align}
\int_{t_1}^{t_2}  \left[ \rho \boldsymbol{u} u_x + \boldsymbol{\Pi}_x^k \right] dt = & \frac{1}{2} \frac{1}{\Delta S_x}  \displaystyle\sum_{i=1}^N  m_i \boldsymbol{\dot{r}}_i  
\left[sgn(x^+- x_{i_2}) - sgn(x_{i_1}- x^+)\right] \Lambda_y(t_k) \tilde{\Lambda}_z(t_k) 
  \nonumber \\
\boldsymbol{\Pi}_x^c = & \frac{1}{4} \frac{1}{ \Delta S_z}\sum_{i,j}^{N} \boldsymbol{f}_{ij}  \left[sgn(x^+- x_j) - sgn(x_i- x^+)\right]] \Lambda_y(\lambda_k) \tilde{\Lambda}_z(\lambda_k).
\label{MOP_implemented}
 \end{align}

Obtaining the expression for crossing on the other flat surfaces $x^-$ and $y^\pm$ follow the process just outlined.
It is complicated in the $z$ direction by the intrinsic surface, which we can evaluate as follows.
The $z$ surface is the last term on the right of \eq{dCV_general},
\begin{align}
 \int_{0}^{1} \Lambda_x \Lambda_y  \frac{\partial \tilde{\Lambda}_z}{\partial s} ds 
& = \int_{0}^{1} \Lambda_x (s) \Lambda_y (s) \frac{\partial}{\partial s}
 \bigg[ H \left( z^+ - \xi(x_\alpha(s), y_\alpha(s)) - z_\alpha(s)   \right) 
\nonumber \\
      & \;\;\;\;\;\;\;\;\;\;\;\; \;\;\;\;\;\;\;\;\;\;\;\;\;\;\;\;\;\;\; - H \left( z^- - \xi(x_\alpha(s), y_\alpha(s)) - z_\alpha(s)   \right) \bigg]  ds 
\nonumber \\
&=   \int_{0}^{1}  \frac{\partial \left(   \xi - z_\alpha \right)}{\partial s} \left[ \delta \left( z^+ - \xi - z_\alpha   \right) -\delta \left( z^- - \xi - z_\alpha   \right) \right] \Lambda_x  \Lambda_y   ds .
\label{crossing_intrinsic}
 \end{align}

Again, using the property of the Delta function to express the sum of the roots of intersection, the crossings of the intrinsic interface by the line of interaction,
\begin{align}
 \delta \left( z^+ - \xi - z_\alpha   \right) = \displaystyle\sum_{k=1}^{N_{roots}} \frac{\delta \left( s - s_{k} \right)}{| \frac{d}{ds} \left( \xi (s_k) - z_\alpha(s_k) \right) |}
 \label{sum_surf}
\end{align}
allows \eq{crossing_intrinsic} to be written as,
\begin{align}
 \int_{0}^{1} \frac{\partial \left(   \xi - z_\alpha \right)}{\partial s}  \delta \left( z^+ - \xi - z_\alpha   \right) \Lambda_x  \Lambda_y ds  &
\nonumber \\
 =  \int_{0}^{1} \frac{\partial}{\partial s}  \left(   \xi(s) - z_\alpha (s) \right)  & \displaystyle\sum_{k=1}^{N_{roots}} \frac{\delta \left( s - s_{k} \right)}{| \frac{\partial}{\partial s} \left( \xi (s_k) - z_\alpha (s_k) \right) |} \Lambda_x (s)  \Lambda_y(s) ds 
\nonumber \\
% =\underbrace{ \frac{\frac{\partial }{\partial s}  \left(   \xi(s_k) - z_\alpha(s_k) \right) }{|\frac{\partial }{\partial s}  \left(   \xi(s_k) - z_\alpha(s_k) \right)|}}_{\textrm{Direction of crossing}}
%& \displaystyle\sum_{k=1}^{N_{roots}} 
% \underbrace{ \left[ H \left( 1- s_{k} \right) - H \left(  - s_{k} \right) \right]  \vphantom{\left( \frac{\partial }{\partial s} \right)} }_{\textrm{Crossing between limits}}  
%\underbrace{ \Lambda_x(s_k)  \Lambda_y(s_k) . \vphantom{\left( \frac{\partial }{\partial s} \right)}}_{\textrm{Crossing on xy CV Surface}} 
 =\underbrace{  \frac{\frac{\partial }{\partial s}  \left(   \xi(s_k) - z_\alpha(s_k) \right)}{|\frac{\partial }{\partial s}  \left(   \xi(s_k) - z_\alpha(s_k) \right)|}  }_{i) \textrm{ Direction of crossing}}
& \displaystyle\sum_{k=1}^{N_{roots}} 
 \underbrace{ \left[ H \left( 1- s_{k} \right) - H \left(  - s_{k} \right) \right]  \vphantom{\left( \frac{\partial }{\partial s} \right)} }_{ii) \textrm{ Crossing between limits}}  
\underbrace{ \Lambda_x(s_k)  \Lambda_y(s_k) . \vphantom{\left( \frac{\partial }{\partial s} \right)}}_{iii) \textrm{ Crossing on xy CV Surface}} .
 \end{align}
The expression is broadly similar to the flat surface case of \eq{flat_surface}, with a more complex expression for the direction of crossing in terms of the intrinsic surface i).
This can be understood using a change of variable,
\begin{align}
\frac{\partial }{\partial s}  \left(   \xi - z_\alpha \right) =\frac{\partial \boldsymbol{r}_\alpha }{\partial s} \cdot \frac{\partial  }{\partial \boldsymbol{r}_\alpha}    \left(   \xi - z_\alpha \right) = \boldsymbol{r}_{12}   \cdot \boldsymbol{\nabla}_\alpha  \left(   \xi - z_\alpha \right) ,
 \end{align} 
 which allows term i) to be written as,
 \begin{align}
 \frac{\frac{\partial }{\partial s}  \left(   \xi - z_\alpha \right)}{|\frac{\partial }{\partial s}  \left(   \xi - z_\alpha\right)|} 
 =  \frac{\boldsymbol{r}_{12}   \cdot \boldsymbol{\nabla}_\alpha  \left(   \xi - z_\alpha \right) }{|\boldsymbol{r}_{12}   \cdot \boldsymbol{\nabla}_\alpha  \left(   \xi - z_\alpha \right) |} 
 \frac{||\boldsymbol{\nabla}_\alpha  \left(   \xi - z_\alpha \right) ||}{||\boldsymbol{\nabla}_\alpha  \left(   \xi - z_\alpha \right) ||} 
 = \frac{\boldsymbol{r}_{12} \cdot \tilde{\boldsymbol{n}}_z  }{ |\boldsymbol{r}_{12} \cdot \boldsymbol{ \tilde{n}_z}|},
  \end{align}
with $||a||$ denoting vector magnitude, which must be positive so can be moved inside the absolute value on the denominator allowing the expression to be written in terms of the normal to the intrinsic surface,
\begin{align}
 \tilde{\boldsymbol{n}}_z \define \frac{ \boldsymbol{\nabla}_\alpha  \left(   \xi - z_\alpha \right)  }{|| \boldsymbol{\nabla}_\alpha  \left(   \xi - z_\alpha \right) ||},
 \end{align}
this can be seen to be in the same form as the flat surface, obtaining the direction of the vector between $\boldsymbol{r}_1$ and $\boldsymbol{r}_2$ dotted with the surface normal.
The crossing between limits of ii) and surface area term iii) are identical in form to the \eq{flat_surface} case, although $s_k$ can no longer be obtained as a closed form solution as in the flat surface case of \eq{MOP}.
The solution of line and surface $z_i + s_k z_{ij} = \xi(x, y)$ to obtain $s_k$ requires the application of a root finding process such as Newton Raphson.

So, the implementation of \eq{Full_compared_stresses} is therefore,
  \begin{align}
\int_{t_1}^{t_2}  \rho \boldsymbol{u} u_z + \boldsymbol{\Pi}_z^k  dt
 = & \frac{1}{\Delta S_z}  \displaystyle\sum_{i=1}^N  m_i \boldsymbol{\dot{r}}_i  
 \frac{\boldsymbol{r}_{i_{12}} \cdot \tilde{\boldsymbol{n}}_z }{|\boldsymbol{r}_{i_{i_{12}}} \cdot \tilde{\boldsymbol{n}}_z |} 
 \displaystyle\sum_{k=1}^{N_{roots}}  \left[ H \left( 1- t_{k} \right) - H \left(  - t_{k} \right) \right] \Lambda_x(t_k)  \Lambda_y(t_k) 
  \nonumber \\
  + & \frac{1}{\Delta S_z} \displaystyle\sum_{i=1}^N  m_i \boldsymbol{\dot{r}}_i  \vartheta_t %sgn\left(\insurfi{+}(x_i(t_2), y_i(t_2), t_2) - \insurfi{+}(x_i(t_2), y_i(t_2), t_1)\right) \Lambda_x(t_2)  \Lambda_y(t_2) 
\nonumber  \\ 
 \boldsymbol{\Pi}_z^c= & \frac{1}{2 \Delta S_z}\sum_{i,j}^{N} \boldsymbol{f}_{ij}  
\frac{\boldsymbol{{r}}_{ij} \cdot \tilde{\boldsymbol{n}}_z }{|\boldsymbol{r}_{ij} \cdot \tilde{\boldsymbol{n}}_z|} \displaystyle\sum_{k=1}^{N_{roots}}  \left[ H \left( 1- \lambda_k \right) - H \left(  - \lambda_k \right) \right] \Lambda_x(\lambda_k)  \Lambda_y(\lambda_k),
 \end{align}
where again the use of $\boldsymbol{r}_{i_{12}}$ to emphasise this is per molecule.
As the surface is no longer flat, the roots $t_k$ and $\lambda_k$ in this expression must be obtained using a form of algorithmic ray tracing, discussed in more detail in the body of the text.
%    \begin{align}
% \int_{t_1}^{t_2}  \rho \boldsymbol{u} u_z + \boldsymbol{\Pi}_z  dt
%  = \frac{1}{\Delta S_z}  \displaystyle\sum_{i=1}^N  m_i \boldsymbol{\dot{r}}_i & 
%  \Big[
%   \frac{\boldsymbol{{r}}_{12} \cdot \boldsymbol{n}_x }{|\boldsymbol{r}_{12} \cdot \boldsymbol{n}_x|}
% + \frac{\boldsymbol{{r}}_{12} \cdot \boldsymbol{n}_y }{|\boldsymbol{r}_{12} \cdot \boldsymbol{n}_y |}
% + \frac{\boldsymbol{{r}}_{12} \cdot \tilde{\boldsymbol{n}}_z }{|\tilde{\boldsymbol{n}}_z \cdot \boldsymbol{{r}}_{12}|}
%  \Big] + \frac{1}{\Delta S_z} \displaystyle\sum_{i=1}^N  m_i \boldsymbol{\dot{r}}_i \vartheta_t
% \nonumber  \\ 
%  + \sum_{n}^{N_\tau} \frac{\Delta t_n}{2 \Delta S_z}\sum_{i,j}^{N} \boldsymbol{f}_{ij} & 
% \Big[  
% \frac{\boldsymbol{{r}}_{ij} \cdot \boldsymbol{n}_x }{|\boldsymbol{r}_{ij} \cdot \boldsymbol{n}_x|}
% + \frac{\boldsymbol{{r}}_{ij} \cdot \boldsymbol{n}_y }{|\boldsymbol{r}_{ij} \cdot \boldsymbol{n}_y|}
% + \frac{\boldsymbol{{r}}_{ij} \cdot \tilde{\boldsymbol{n}}_z }{|\boldsymbol{r}_{ij} \cdot \tilde{\boldsymbol{n}}_z|}
% \Big] 
%  \end{align}

\section{Linear Algebra}
\label{sec:LA}

The least square minimisation of the weighting function to fit the intrinsic surface has been discussed in the literature, most extensively in the supplementary material of \citet{Longford_et_al18}. 
However, a slightly different notation is given here to clarify a few steps, starting from,
\begin{align}
W( a_{\mu \nu} ) = \frac{1}{2}\displaystyle\sum_{p=1}^{N_p} \left[ z_p - \displaystyle\sum_{\mu=-k_u}^{k_u} \displaystyle\sum_{\nu=-k_u}^{k_u}  a_{\mu \nu}(t) f_\mu (x_p) f_\nu(y_p) \right]^2 +  \psi \tilde{A}.
\end{align}
This can be seen to be a linear algebra problem by defining the vector $\boldsymbol{f}(x,y)$ where each row is a wavevector,
\begin{align}
\boldsymbol{f} (x,y) =
\begin{bmatrix}
f_{-k_u} (x) f_{-k_u}(y)   \\
f_{-k_u} (x) f_{-k_u+1}(y)    \\
\dots \\
f_{-k_u} (x) f_{k_u}(y) \\
f_{-k_u+1} (x) f_{-k_u}(y) \\
\dots \\
f_{k_u} (x) f_{k_u}(y)
\end{bmatrix} 
\end{align}
so the total matrix $\boldsymbol{F}$ can be defined by stacking matrices $\boldsymbol{f}^T$ for all pivot locations,
\begin{align}
\boldsymbol{F} =
\begin{bmatrix}
\boldsymbol{f}^T (x_1,y_1)  \\
\boldsymbol{f}^T (x_2,y_2)   \\
\dots \\
\boldsymbol{f}^T (x_{N_p},y_{N_p})
\end{bmatrix} 
\end{align}
and defining $\boldsymbol{z} = [z_1, z_2, ..., z_{N_p}]^T$, $\boldsymbol{a} = a_{\nu \mu}$ and $\tilde{A} = \boldsymbol{a}^T \boldsymbol{a} \textbf{B}$ with $  \textbf{B}= 4 \pi^2 \textrm{ diag}(\nu^2 + \mu^2 )$ a matrix with values only on the diagonal and zero elsewhere. 
%$\tilde{A} = 4 \pi^2 a_{\nu \mu} a_{\nu' \mu'} \left[\nu^2 + \mu^2 \right]$
% so $\xi = \boldsymbol{F}\boldsymbol{a}$ and area can be defined as,
% \begin{align}
% \tilde{A} = A + \frac{1}{2} \int \int \left( \frac{\partial \xi}{\partial x} \right)^2 +\left( \frac{\partial \xi}{\partial y} \right)^2  dx dy 
% = A + \frac{1}{2} \int \int \left( \boldsymbol{F}_x \boldsymbol{a} \right)^2 +\left( \boldsymbol{F}_y \boldsymbol{a}\right)^2  dx dy 
% \nonumber \\
% = A + \frac{1}{2} \int \int \boldsymbol{F}_x^T \boldsymbol{a}^T  \boldsymbol{F}_x \boldsymbol{a} +\boldsymbol{F}_y^T \boldsymbol{a}^T \boldsymbol{F}_y \boldsymbol{a}  dx dy 
% \end{align}
% 
% $\tilde{A} = 4 \pi^2 a_{\nu \mu} a_{\nu' \mu'} \left[\nu^2 + \mu^2 \right]$
% and area $\tilde{\boldsymbol{A}} =  4 \pi^2 \Psi(\nu, \mu) a_{\nu \mu} \left[\nu^2 L_x/L_y + \mu^2 L_y/L_x \right]$ with $\Psi(\nu, \mu)$ defined in \citet{Longford_et_al18} to allow domains which are not square, with $\Psi(\nu=0, \mu=0) = 4$, $\Psi(\nu \ne 0, \mu = 0)=\Psi(\nu=0, \mu \ne 0) = 2$ and $\Psi(\nu \ne 0, \mu \ne 0) = 1$.
So, $W$ in \eq{Minimise} becomes,
\begin{align}
W(\boldsymbol{a}) =  \frac{1}{2} || \boldsymbol{z} - \boldsymbol{F} \boldsymbol{a} ||^2 +  \psi \tilde{\boldsymbol{A}},
\end{align}
and the minimum is ontained by setting the derivative with respect to $\boldsymbol{a}$ to zero,
\begin{align}
\boldsymbol{\nabla}_a W =  - \boldsymbol{F}^T \left( \boldsymbol{z} - \boldsymbol{F} \boldsymbol{a} \right) +  \psi \boldsymbol{\nabla}_a\tilde{\boldsymbol{A}} = 0,
\end{align}
and $\boldsymbol{\nabla}_a\tilde{\boldsymbol{A}} = \textbf{B} \boldsymbol{a} $ so the optimal value of $\boldsymbol{a}$ is obtained by solving this equation,
\begin{align}
\boldsymbol{a} =  ( \boldsymbol{F}^T \boldsymbol{F} - \psi\textbf{B})^{-1} \boldsymbol{F}^T \boldsymbol{z} .
\end{align}
To maximise efficiency, LAPACK \citep{Lapack} is used to solve this equation.
In practice this means the matrix $\boldsymbol{F}$ of size $M \times N_p$ with $M = 4k_u^2$ wavenumbers and $N_p$ pivot positions, is multiplied by its transpose (using e.g. Lapack DGEMM).
The constraint is then applied by subtracting a matrix of size $M \times M$ with just diagonal elements $\boldsymbol{B} = 4 \pi^2 \psi \left[\mu^2 + \nu^2\right]$ that are non-zero.
This can then be used in a linear algebra solver (e.g. Lapack DGESV) with right-hand side $\boldsymbol{F}^T \boldsymbol{z}$ to get the values of $\boldsymbol{a}$.

\section{A Derivation of the Volume Average Form}
\label{sec:surface_terms}

In this section, we consider how to obtain the volume average (VA) expressions from \eq{Full_time_derivative_momentum},
\begin{align}
\frac{d}{dt} \int_V \rho \boldsymbol{u} dV = \frac{d}{dt} \displaystyle\sum_{i=1}^N m_i \dot{\boldsymbol{r}}_i \vartheta_i 
= \displaystyle\sum_{i=1}^N m_i \dot{\boldsymbol{r}}_i \frac{d \vartheta_i}{dt} + \displaystyle\sum_{i=1}^N m_i \ddot{\boldsymbol{r}}_i \vartheta_i ,
\end{align}
where the continuum left-hand side is expressed using the divergence theorem,
\begin{align}
\frac{d}{dt} \int_V \rho \boldsymbol{u} dV = -\oint_S  \left[ \rho \boldsymbol{u} \boldsymbol{u} + \boldsymbol{\Pi} \right] \cdot d\textbf{S}
=- \int_V \frac{\partial}{\partial \boldsymbol{r}} \cdot \left[ \rho \boldsymbol{u} \boldsymbol{u} + \boldsymbol{\Pi} \right] dV,
\end{align}
while the right-hand side is expanded  using \eq{dsurfmass_z}, \eq{Force_term} and \eq{derivative_lambda},
\begin{align}
\frac{d}{dt} \displaystyle\sum_{i=1}^N m_i \dot{\boldsymbol{r}}_i \vartheta_i = \displaystyle\sum_{i=1}^N m_i \dot{\boldsymbol{r}}_i \left( \dot{\boldsymbol{r}}_i \cdot \frac{\partial \vartheta_i }{\partial \boldsymbol{r}_i} + \frac{\partial \xi }{\partial t} \frac{\partial \vartheta_i }{\partial \xi} \right) + \displaystyle\sum_{i,j}^N \boldsymbol{f}_{ij} \int_0^1\boldsymbol{r}_{ij} \cdot \frac{\partial \vartheta_\lambda }{\partial \boldsymbol{r}_\lambda} d \lambda,
\end{align}
by assuming $\partial \vartheta_i / \partial \boldsymbol{r}_i = -\partial \vartheta_i / \partial \boldsymbol{r}$, $\partial \vartheta_{\lambda} / \partial \boldsymbol{r}_\lambda = -\partial \vartheta_\lambda / \partial \boldsymbol{r} $ and $ \partial \xi /\partial t=0$, we can express everything as a derivative in terms of $\boldsymbol{r}$,
\begin{align}
\int_V \frac{\partial}{\partial \boldsymbol{r}} \cdot \left[ \rho \boldsymbol{u} \boldsymbol{u} + \boldsymbol{\Pi} \right] dV 
=
\frac{\partial}{\partial \boldsymbol{r}}  \cdot \left[ \displaystyle\sum_{i=1}^N m_i \dot{\boldsymbol{r}}_i \dot{\boldsymbol{r}}_i  \vartheta_i  + \displaystyle\sum_{i,j}^N \boldsymbol{f}_{ij} \boldsymbol{r}_{ij} \cdot \int_0^1 \vartheta_\lambda d \lambda \right],
\end{align}
and so, comparing the expressions inside the derivative and assuming an average value for the volume yields the VA pressure given in previous work,
\begin{align}
 \rho \boldsymbol{u} \boldsymbol{u} + \PressureVA  = \frac{1}{\Delta V} \left[  \displaystyle\sum_{i=1}^N  m_i \dot{\boldsymbol{r}}_i  \dot{\boldsymbol{r}}_{i}  \vartheta_i +  \frac{1}{2}\displaystyle\sum_{i,j}^N \boldsymbol{f}_{ij} \boldsymbol{r}_{ij} \int_0^1 \vartheta_\lambda d \lambda \right].
\end{align}
%A more general derivation of these volume average expression would follow from the time evolution of \eq{Full_time_derivative_momentum}, which are shown in the appendix \ref{sec:surface_terms} to give additional surface curvature dependant terms.

%As shown in the appendix \ref{sec:surface_terms}, this form of the volume average omit surface curvature dependant terms.
%\displaystyle\sum_{i=1}^N m_i \dot{\boldsymbol{r}}_i \frac{d \vartheta_i}{dt} = \displaystyle\sum_{i=1}^N m_i \dot{\boldsymbol{r}}_i \frac{\partial \vartheta_i}{\partial t} + \frac{\partial \vartheta_i}{\partial \boldsymbol{r}_i} 

However, the assumptions $\partial \vartheta_{\alpha} / \partial \boldsymbol{r}_{\alpha} = -\partial \vartheta_{\alpha} / \partial \boldsymbol{r}$ and $ \partial \xi /\partial t=0$ are not valid, as will be shown here for the configurational term, resulting in an extra term in the VA expression for a volume between curved interfaces.
Pressure is defined in the \citet{Irving_Kirkwood} process by collecting terms inside the gradient with respect to $\boldsymbol{r}$ and comparing forms to the continuum expression $\partial / \partial \boldsymbol{r} \cdot \boldsymbol{\Pi} $.
In the pointwise \citet{Irving_Kirkwood}, this uses the property of the Dirac delta $ \partial / \partial \boldsymbol{r}_\lambda \delta(\boldsymbol{r}-\boldsymbol{r}_{\lambda}) = -\partial / \partial \boldsymbol{r} \delta(\boldsymbol{r}-\boldsymbol{r}_{\lambda})  $.
For an integrated control volume between intrinsic surfaces, to see if the same process can be applied, we compare the derivatives,
$\frac{\partial \vartheta_\lambda }{\partial \boldsymbol{r}_\lambda}$ and $\frac{\partial \vartheta_\lambda }{\partial \boldsymbol{r}}$.
Starting with the derivative of $\vartheta_\lambda$ with respect to $\boldsymbol{r}$,
\begin{align} 
\frac{\partial \vartheta_\lambda }{\partial \boldsymbol{r}} = 
	 \boldsymbol{i}\frac{\partial \Lambda_x}{\partial x} \Lambda_y \tilde{\Lambda}_z 
+    \boldsymbol{j}\Lambda_x  \frac{\partial \Lambda_y}{\partial y}  \tilde{\Lambda}_z
+  \boldsymbol{k}\Lambda_x \Lambda_y \frac{\partial  \tilde{\Lambda}_z }{\partial z},
\end{align}
where e.g. $\frac{\partial \Lambda_x}{\partial x} = \delta(x^+-x_\lambda) - \delta(x^- -x_\lambda)$ and  $\frac{\partial \tilde{\Lambda}_z}{\partial z} = \delta(\insurfl{+}-z_\lambda) - \delta(\insurfl{$-$} -z_\lambda)$ as $\insurfl{$\pm$} = z \pm \Delta z + \xi$ so $d \insurfl{$\pm$} / dz = 1$.

Next, consider the derivative of $\vartheta_\lambda$ with respect to $\boldsymbol{r}_\lambda$
\begin{align} 
\frac{\partial \vartheta_\lambda }{\partial \boldsymbol{r}_\lambda} = 
	 \boldsymbol{i} \frac{\partial \Lambda_x}{\partial x_\lambda} \Lambda_y \tilde{\Lambda}_z 
+    \boldsymbol{j}  \Lambda_x  \frac{\partial \Lambda_y}{\partial y_\lambda}  \tilde{\Lambda}_z
+  \Lambda_x \Lambda_y \left[\boldsymbol{i} \frac{\partial  \tilde{\Lambda}_z }{\partial x_\lambda} + \boldsymbol{j} \frac{\partial  \tilde{\Lambda}_z }{\partial y_\lambda} + \boldsymbol{k}\frac{\partial  \tilde{\Lambda}_z }{\partial z_\lambda} \right] .
\end{align}
Noting that $\frac{\partial \Lambda_x}{\partial x_\lambda} = - \left[\delta(x^+-x_\lambda) -\delta(x^- -x_\lambda) \right] = -\frac{\partial \Lambda_x}{\partial x}$ and similar for $y$, while the derivative of $\tilde{\Lambda}_z$ with respet to $z$ is $\frac{\partial \tilde{\Lambda}_z}{\partial z_\lambda} = \delta(\insurfl{+}-z_\lambda) - \delta(\insurfl{$-$} -z_\lambda)$, so $\frac{\partial \tilde{\Lambda}_z}{\partial z} = -\frac{\partial \tilde{\Lambda}_z}{\partial z_\lambda}$.
Using these equivalences,
\begin{align} 
\frac{\partial \vartheta_\lambda }{\partial \boldsymbol{r}_\lambda} =  
-\boldsymbol{i} \frac{\partial \Lambda_x}{\partial x} \Lambda_y \tilde{\Lambda}_z 
-\boldsymbol{j} \frac{\partial \Lambda_y}{\partial y} \Lambda_x \tilde{\Lambda}_z
-\boldsymbol{k} \frac{\partial \tilde{\Lambda}_z }{\partial z}  \Lambda_x \Lambda_y 
+  \Lambda_x \Lambda_y \left[\boldsymbol{i} \frac{\partial  \tilde{\Lambda}_z }{\partial x_\lambda} + \boldsymbol{j} \frac{\partial  \tilde{\Lambda}_z }{\partial y_\lambda} \right] 
\nonumber \\
=-\frac{\partial \vartheta_\lambda }{\partial \boldsymbol{r}} + 
\left[ \boldsymbol{i} \left( \frac{\partial \insurfl{+} }{\partial x_\lambda} dS_{z\lambda}^+ - \frac{\partial \insurfl{$-$} }{\partial x_\lambda} dS_{z\lambda}^-\right) + \boldsymbol{j} \left(\frac{\partial \insurfl{+} }{\partial y_\lambda} dS_{z\lambda}^+  - \frac{\partial \insurfl{$-$} }{\partial y_\lambda} dS_{z\lambda}^- \right) \right],
\end{align}
where 
\begin{align} 
\frac{\partial  \tilde{\Lambda}_z }{\partial x_\lambda} =  \left[ \frac{\partial \insurfl{+} }{\partial x_\lambda} \delta(\insurfl{+} - z_\lambda) - \frac{\partial  \insurfl{$-$} }{\partial x_\lambda}  \delta(\insurfl{$-$} - z_\lambda) \right],
\end{align}
and similar for the $y$ derivative, with the surface notation used, $dS_{z\lambda}^\pm = \delta(\insurfl{$\pm$} - z_\lambda) \Lambda_x \Lambda_y$.
The full expression is therefore,
\begin{align} 
\displaystyle\sum_{i,j}^N \boldsymbol{f}_{ij} \left[\vartheta_i - \vartheta_j \right] &= \displaystyle\sum_{i,j}^N \boldsymbol{f}_{ij} \boldsymbol{r}_{ij} \cdot 
\int_0^1    \frac{\partial \vartheta_\lambda }{\partial \boldsymbol{r}_\lambda}d \lambda
 \nonumber \\
  &= -\frac{\partial }{\partial \boldsymbol{r}} \cdot \underbrace{\displaystyle\sum_{i,j}^N \boldsymbol{f}_{ij} \boldsymbol{r}_{ij}  \int_0^1   \vartheta_\lambda d \lambda }_{\PressureVA {}^c  }
 + \underbrace{ \displaystyle\sum_{i,j}^N \boldsymbol{f}_{ij} \boldsymbol{r}_{ij} \cdot \int_0^1 \left[\frac{\partial \insurfl{+} }{\partial \boldsymbol{r}_\lambda}  dS_{z\lambda}^+ - \frac{\partial \insurfl{$-$} }{\partial \boldsymbol{r}_\lambda} dS_{z\lambda}^- \right]   d\lambda }_{\textrm{Extra Term} },
\end{align}
using $\partial \insurfl{$\pm$} / \partial z_\lambda = 0$ to write the extra term in vector form. 

\bibliographystyle{unsrtnat}
\bibliography{ref.bib}

\end{document}